# Comparison of various UHMWPE formulations from contemporary total knee replacements before and after accelerated aging

Petr Fulin[1], Veronika Gajdosova[2], Ivana Sloufova[3], Jiri Hodan[2], David Pokorny[1], Miroslav Slouf[2]*

[1] 1st Department of Orthopaedics, First Faculty of Medicine of Charles University and Motol University Hospital, V Uvalu 84, 15006 Prague, Czech Republic

[2] Institute of Macromolecular Chemistry, Czech Academy of Sciences, Heyrovskeho nam. 2, 16206 Praha 6, Czech Republic

[3] Charles University, Faculty of Science, Department of Physical and Macromolecular Chemistry, Hlavova 2030, 128 40 Prague 2, Czech Republic

* Corresponding author e-mail: slouf@imc.cas.cz

## Abstract

We have collected 21 different formulations of ultrahigh molecular weight polyethylene (UHMWPE), which have been employed as liners in contemporary total knee replacements (TKR). The UHMWPE liners were bought from the most important manufacturers on the orthopedic market in the Czech Republic as of 2020. The collected liners represented a broad range of both traditional and modern UHMWPE formulations, which differed by the level of crosslinking, type of thermal treatment, sterilization and/or stabilization. All obtained UHMWPE's were characterized by multiple methods immediately after purchase and after the accelerated aging in $H_2O_2$. The experimental results (oxidative degradation, structure changes, and micromechanical properties) were correlated with manufacturer's data (crosslinking, thermal treatment, sterilization, and stabilization). The investigated UHMWPE liners exhibited significant differences in their properties, namely in their resistance to long term oxidative degradation. The stiffness-related mechanical properties showed a strong correlation with the overall crystallinity. The crystallinity depended mostly on the oxidative degradation of the UHMWPE liners, while the thermal treatment played a minor role. The highest resistance to oxidation and wear, which promises the best *in vivo* performance, was found for the crosslinked UHMWPE formulations with biocompatible stabilizers (such as α-tocopherol, which is the key component of vitamin E).

## 1 Introduction

Total joint replacements (TJR) have become an essential part of modern orthopedics [1,2].Ultra-high molecular weight polyethylene (UHMWPE) is used as the load-bearing component for the total joint replacements (TJR) since 1960´s due to its biocompatibility, good friction properties, and sufficient mechanical performance [3]. The first UHMWPE formulations for TJR were made from virgin polymer and sterilized by gamma irradiation. Nevertheless, the gamma-sterilized UHMWPEs showed numerous early failures after ca 10 years [3,4]. It was shown that the UHMWPEs subjected to ionizing radiation (such as gamma rays or electron beam) contained residual radicals from the irradiation, which could survive in the polymer for many years, causing long-term oxidative degradation, followed by the deterioration of mechanical properties including wear resistance [5–7]. The oxidation (i.e. the long-term oxidative degradation) and wear (i.e. the release of micro- and nanoscale particles from the articulating polymer surface) were shown to be the most important material-related reasons of the TJR failures, despite some biomechanics-related differences among hips, knees and other joints [8].

The abovementioned limitations of the gamma-sterilized UHMWPEs resulted in the development of modern, crosslinked UHMWPE formulations with higher oxidation and wear resistance. The 1st generation crosslinked UHMWPEs were introduced in late 1990's and early 2000's [9]. They have been prepared in three steps comprising (i) crosslinking by ionizing radiation to increase wear resistance, (ii) thermal treatment to eliminate the residual radicals from the crosslinking, and (iii) the obligatory final sterilization step [10,11]. The thermal treatment comprises either *remelting* (heating above the polymer melting temperature, with a slightly negative impact on mechanical performance and elimination of all residual radicals) or *annealing* (the heating below the melting temperature with lower impact on mechanical performance, but an incomplete elimination of the residual radicals) [7,12,13]. The sterilization by gamma irradiation has been replaced by ethylene oxide or gas plasma sterilization by many manufacturers in order not to introduce additional residual radicals [14,15]. The 2nd generation crosslinked UHMWPEs started to be developed in late 2000's, aimed at the preservation of the wear resistance of the 1st generation and further increase in the oxidation stability and mechanical performance. This is usually achieved by addition of biocompatible stabilizers [16,17], although some alternatives, such a sequential irradiation [18], e-beam irradiation at elevated temperatures [19], peroxide crosslinking [20,21], and morphology tuning using high temperatures of pressures [8,22] were tested as well. The biocompatible stabilizers scavenge the residual radicals, which improves the oxidation resistance, enables to omit the remelting step and, as a result, it can improve the final mechanical performance. In the field of TJR, the most widely used stabilizer is vitamin E (or, more precisely, α-tocopherol, which is the most active component of the natural vitamin). Alternative biocompatible stabilizers, such as Covernox™ which was introduced by DePuy company in their AOX™ polyethylene for TJR, can be applied as well on condition they are approved by relevant authorities [23]; it is worth noting that the Covernox™ stabilizer, whose chemical name is *pentaerythritol tetrakis[(3,5-di-tert-butyl-4-hydroxyphenyl)propionate]*, has the same structural formula as Irganox 1010®, a common commercial stabilizer for technical applications [24].

The crosslinked UHMWPEs predominate in the field of total hip replacements (TJR), while in the field of total knee replacements (TKR) the situation is not so clear [8,17]. The main reason consists in the different biomechanical characteristics of the hip and knee joints. In TKR, the wear mechanisms are different, the polymer insert is subjected to greater contact stresses, and the mechanical properties of UHMWPE are believed to play a more important role [25]. As the 1st generation UHMWPEs exhibit higher wear resistance at the expense of mechanical performance, namely the fatigue properties and toughness [7], the application of crosslinked polyethylene for TKR was delayed compared to THR. The 2nd generation crosslinked UHMWPEs showed better results in laboratory studies than conventional UHMWPE mostly due to the fact that the stabilizers mitigated the oxidative degradation significantly [17,25], but the clinical performance of crosslinked UHMWPEs in TKR has not been so convincing. The registry study of Boyer et al. [26] concluded that the crosslinking seems to have a low, if any, effect on knee arthroplasty survival. The recent systematic review of Bistolfi et al. [27], based on the data from six clinical studies, showed just statistically insignificant differences between conventional and crosslinked UHMWPEs in TKR regarding clinical, radiological, and functional outcomes; the superiority of crosslinked UHMWPE over conventional UHMWPE in TKR has not been proved.

In this contribution, we collected new UHMWPE liners from 21 different TKRs. They represented the most frequently implanted TKRs in the Czech Republic in years 2020-2021; the Czech Republic is a country with open market where most of the worldwide TKR manufacturers offer their products as documented in Table 1. All liners were characterized by six different techniques (IR microspectroscopy, microindentation hardness testing, differential scanning calorimetry, thermogravimetric analysis, solubility testing, microscopy) before and after accelerated aging. The first objective was to compare how much the current UHMWPE formulations for TKR differ in their structure and properties. The second and more practical objective was to provide an unbiased, non-commercial, experiment-based set of recommendations to the orthopedic community, which of the current UHMWPE types promise the longest lifetime *in vivo* from the materials science perspective.

# 2 Experimental

## 2.1 Materials

We collected 21 new UHMWPE liners (Table 1) of various TKR. All UHMWPEs were delivered to a hospital (the workplace of the 1st author of this work) in a standard way, in the original packaging as if they should be used for the implantation. The information about the UHMWPE liners (the UHMWPE type, modifications, crosslinking, thermal treatment, and sterilization) was obtained from the package leaflet, product sheets of individual manufacturers, and/or from the information available in the literature [28,29]. The key pieces of information are collected in Table 1, additional information is available at request to the corresponding author of this study (MS).

Table 1

## 2.2 Sample preparation

All 21 UHMWPE liners were cut to 2 mm thick plates perpendicular to the articulating surfaces. The cutting was performed at room temperature, slowly, with a sharp blade, in order to avoid overheating and structure changes as described elsewhere [30]. The 2 mm plates were the starting samples for both accelerated aging (section 2.3) and all characterization methods (section 2.4). For some methods, the 2 mm plates were further adjusted by cutting or microtomy, as described below.

## 2.3 Accelerated aging

The accelerated aging of UHMWPE total knee replacements was carried out in the 0.1 M aqueous solution of hydrogen peroxide ($H_2O_2$) at 70 °C for 30 days. Importantly, the experimental vessel with $H_2O_2$ solution was made of glass without any metal components, in order to avoid the decomposition of hydrogen peroxide, which is very sensitive to metal ions. Small pieces of UHMWPE liners (ca 10 x 10 x 2 mm) were labelled using color glass beads strung on a glass string (i.e. the whole experiment was performed with resistant and strictly non-metallic components). The beads acted both as a sample identifier and dead weight, forcing the light polyethylene plates to be fully submerged at the bottom of the experimental vessel. The whole system was heated to 70 °C in a standard laboratory oven with a temperature controller. The aging medium was replaced with a fresh one every 7 days, in order to keep the $H_2O_2$ concentration more constant and reproducible. The accelerated aging was carried out for 30 days. Additional information about our accelerated aging method can be found in section 1 of the Supplementary information file.

## 2.4 Characterization methods

All 21 UHMWPE liners (Table 1) were characterized by eight testing methods: infrared microspectroscopy (IR), instrumented microindentation hardness testing (MHI), non-instrumented microindentation hardness testing (MH), differential scanning calorimetry (DSC), thermogravimetric analysis (TGA), determination of the degree of crosslinking (solubility measurements, SOL), light microscopy (LM), and scanning electron microscopy (SEM). The key methods for characterization of UHMWPE structure (IR) and properties (MHI and MH) were applied to both new and aged samples. Additional methods for crystallinity verification (DSC), testing of stabilizer presence (TGA), and estimation of crosslinking (SOL) were applied only to the original samples before aging. Microscopic methods (SEM, LM) were employed in the characterization of the surface damage after the accelerated aging. We could not employ standard macroscale characterization methods, such as tensile testing, due to the limited size and irregular shape of the UHMWPE liners.

### 2.4.1 Infrared microspectroscopy

Infrared spectra (IR) were measured using an IR microscope (Thermo Nicolet 6700 with an FTIR microscope Continuum, equipped with the MCT detector). The 2 mm plates were cut into the thin slices ~200 µm using a sledge microtome (Meopta; Czech Republic). The infrared spectra of the microtomed slices were measured with the IR microscope in the transmission mode using accumulation of 16 scans with a resolution of 4 cm$^{-1}$. The spectra were measured as linear scans, i.e. in a line along the direction perpendicular to the surface of the plaque. The distance between the individual measurements in the line was 100 µm. From the IR measurement, several IR indexes were calculated from each spectrum: The oxidation index (OI; proportional to the local oxidative degradation; Eq. 1 [31,32]), *trans*-vinylene index (VI; proportional to the absorbed radiation dose; Eq. 2 [30,33]), and crystallinity index (CI; proportional to the local volume fraction of the crystalline phase; Eq. 3 [34,35]).

$$\mathrm{OI} = \frac{A_{1720}}{A_{1370}} \quad (1)$$

$$\mathrm{VI} = \frac{A_{960}}{A_{1370}} \quad (2)$$

$$\mathrm{CI} = \frac{A_{1897}/A_{1303}}{A_{1897}/A_{1303} + 1} \quad (3)$$

For the original, non-aged samples, the values of OI, CI and VI were averaged through the whole line scan (profile). For the aged samples, we calculated two values for OI and CI indexes: the maximum value of the index close to the surface (referred as OI(max) and CI(max)) and the average value of the index in the central region (referred as OI(ave) and CI(ave)). The VI indexes did not change as a function of distance from the exposed surface and, as a result, we determined just their average values (VI(ave)).

### 2.4.2 Differential scanning calorimetry

Differential scanning calorimetry (DSC) was performed using Q2000 calorimeter (TA Instruments, USA). The measurements were carried out in the heating-cooling-heating cycles from 0 °C to +180 °C at the heating rate of 10 °C/min and under the nitrogen purge of 50 mL/min. The melting temperature ($T_m$) and crystallization temperature ($T_c$) were determined from the position of the melting peak maximum on heating curve and crystallization peak maximum on cooling curve, respectively. Melting and crystallization onset temperatures were obtained as intersection of baseline and inflection line of the melting and crystallization peak, respectively. Crystallinity (CR) was calculated as the area under the melting peak ($\Delta H_m$) normalized to 100 % crystalline polyethylene ($\Delta H_{\mathrm{m}}^{0}$ = 293.6 J/g; Eq. 4; [31]). All samples were measured three-times and the results were averaged.

$$\mathrm{CR} = \Delta H_m / \Delta H_m^0 \qquad (4)$$

### 2.4.3 Thermogravimetric analysis

Thermal gravimetric analysis (TGA) was performed using thermogravimetric analyzer Perkin Elmer TGA 7 apparatus equipped with Pyris 1 software at the heating rate of 10 °C/min, under 20 mL/min air flow, and within temperature range of 35–400 °C. The TGA curves were used for the determination of oxidation onset temperature (OOT). The OOT value was estimated from the intersection of the baseline and the tangent drawn to the initial short weight gain in the TGA curve (due to the absorption of oxygen by the sample), which was followed by the steep weight loss (due to formation of volatile decomposition products of the sample).

### 2.4.4 Micromechanical properties

Micromechanical properties were measured using two devices: a non-instrumented microindentation hardness tester (VMHT Auto Man; Uhl, Germany) and an instrumented microindentation hardness tester (MCT tester; CSM, Switzerland). Both non-instrumented microindentation (MH) and instrumented microindentation (MHI) measurements were carried out using Vickers method: a diamond square pyramid (with the angle between the non-adjacent faces 136°) was forced against the flat surface of a specimen. The flat smooth surfaces for the MH and MHI testing were prepared by cutting from the 2 mm UHMWPE plates with a rotary microtome RM 2155 (Leica, Vienna, Austria). From each sample, three independent cut surfaces were prepared. For each measured surface, at least 10 independent measurements/indentations were made and the final results were averaged.

In the case of MH, we calculated Vickers microhardness ($H_v$) from the loading force ($F$) and average diagonal length of the indent ($d$) according to Eq. 5. We used the maximal loading force $F$ = 50 gf = 490.5 mN, immediate loading and unloading, and the dwell time (i.e. time of maximal load) = 60 s.

$$H_V = 1.854\, F/d^2 \qquad (5)$$

In the case of MHI, the device records the loading force ($F$) as a function of time ($t$) and penetration depth ($h$). The MHI measurements parameters were set as similar as possible to the non-instrumented MH measurements: loading force 500 mN (~ 50 gf), dwell time 60 s, and linear loading and unloading rates 15 000 mN/min (~2s). The properties evaluated from MHI measurements were indentation modulus ($E_{IT}$, proportional to macroscopic elastic modulus; Eq. 6), indentation hardness ($H_{IT}$; proportional to macroscopic yield stress, Eq. 7), Martens hardness ($H_M$, also referred as universal hardness and proportional to $H_{IT}$, Eq. 8), indentation creep ($C_{IT}$, related to the macroscopic creep, Eq.

9), and elastic part of the indentation work ($\eta_{IT}$, defined as ratio of elastic deformation to total deformation, Eq. 10):

$$E_{IT} \propto S/\sqrt{A_p}(h_c) \tag{6}$$

$$H_{IT} = F_{max}/A_p(h_c) \tag{7}$$

$$H_M = F_{max}/A_d(h_2) \tag{8}$$

$$C_{IT} = (h_2 - h_1)/h_1 \cdot 100\% \tag{9}$$

$$\eta_{IT} = W_{elast}/W_{total} \cdot 100\% \tag{10}$$

In Eq. 6–10, $S$ is stiffness (slope at the beginning of the unloading curve), $h_c$ is the contact depth calculated by the control software according to the theory of Oliver and Pharr (O&P theory; [36]), $A_p$ and $A_d$ are the projected and developed area of the indent, respectively, $h_1$ and $h_2$ are the penetration depths at the beginning and at the end of maximal loading, respectively, $W_{elast}$ is the area under the unloading curve and $W_{total}$ is the area under the loading curve [37]. More detailed description of all formulas can be found elsewhere [38]. The calculations of $E_{IT}$ and $H_{IT}$, are based on the theory O&P theory [36], while the calculation of $H_V$, $H_M$, $C_{IT}$ and $\eta_{IT}$ are independent on the O&P theory [39]. The exact definitions of above-listed micromechanical properties can be found in suitable reviews or textbooks dealing with micro- and/or nanoindentation [36,40,41]; a more detailed description of the MHI experiments also in our recent studies [39,42].

### 2.4.5 Electron spin resonance

Electron spin resonance (ESR) was used for the detection of the possible residual radicals after UHMWPE modifications. The ESR spectra were measured for all samples when all other experiments had been finished in order to verify if the residual radicals are still present. The measurements were performed with a commercial ESR spectrometer (Bruker ELEXSYS E-540; Bruker, Germany) as described previously [43]. The concentrations of residual radicals were estimated from the double integral of the spectra, which is proportional to the number of detected spins.

### 2.4.6 Swelling experiments

Equilibrium swelling experiments were performed to estimate UHMWPE crosslinking density as described elsewhere [43,44]. Briefly, small pieces of TKR samples (ca 1x1x0.2 cm; ~0.5 g) were submerged into xylene (ca 100 mL) with the addition of 1% of commercial antioxidant BHT (butylated hydroxytoluene) and the xylene solution was boiled under reflux (138 °C) for 8 hours. The initial mass of sample ($m_0$) was determined before each experiment. Mass of the sample swollen with xylene ($m_x$) was measured after taking out the specimen from the xylene. The mass of the dried sample $m_D$ was determined after the swollen sample was dried to the constant weight. The basic parameters from swelling experiments, which are the insoluble fraction ($g$, Eq. 11), soluble fraction ($e$, Eq. 12) and equilibrium mass swell ratio ($q_m$, Eq. 13), were calculated according to our previous study [43]:

$$g = \frac{m_D}{m_0} \times 100\% \tag{11}$$

$$e = \frac{m_0 - m_D}{m_0} \times 100\% \tag{12}$$

$$q_m = \frac{m_X}{m_D} \quad (13)$$

Crosslinking density was estimated from the Flory network theory [45] as described by Abreau et al. [44]. The procedure comprises the calculation of equilibrium volume swell ratio ($q_V$, Eq. 14), crosslink density ($N_e$, Eq. 15; moles of crosslinks per dm$^3$), and molecular weight between crosslinks ($M_e$, Eq. 16; in g/mol):

$$q_V = \frac{V_{xyl} + V_0}{V_0} \quad (14)$$

$$N_e = -\frac{\ln(1 - q_V^{-1}) + q_V^{-1} + \chi q_V^{-2}}{\phi_1 (q_V^{-1/3} - q_V^{-1}/2)} \quad (15)$$

$$M_e = \frac{1}{\nu N_e} \quad (16)$$

In Equations 14–16 above, $V_{xyl}$ is the volume of absorbed xylene, $V_0$ is the initial volume of the sample, $\chi = 0.33 + 0.55 / q_V$ is the Flory interaction parameter, $\phi_1 = 0.136$ dm$^3$/mol is the molar volume of the solvent, and $v$ is the specific volume of UHMWPE in xylene at 130 °C with $v^{-1} = 920$ g/dm$^3$. $V_{xyl}$ and $V_0$ could be calculated easily from the measured masses ($m_0$, $m_X$, and $m_D$) and known densities of UHMWPE ($\rho_{PE} = 0.94$ g/cm$^3$) and xylene ($\rho_{xyl} = 0.75$ g/cm$^3$) at 130 °C using the following relations: $V_{xyl} = (m_X - m_0)/\rho_{xyl}$ and $V_0 = m_0/\rho_{PE}$.

### 2.4.7 Light microscopy and scanning electron microscopy

Light microscopy (LM) and scanning electron microscopy (SEM) were employed to observe morphology changes and/or imprints after microindentation on UHMWPE surfaces before and after accelerated aging. LM micrographs were obtained with a light microscope Nikon Eclipse 80i (Nikon, Japan) equipped with a digital camera ProgRes CT3 (Jenoptik, Germany). The surfaces were observed in reflected light. SEM micrographs were obtained with a scanning electron microscope MAIA3 (Tescan, Czech Republic). To minimize electron beam damage, the samples were sputter-coated with platinum (vacuum sputter coater SCD 050; Leica, Austria; thickness of the Pt layer ca 4 nm). The sputter-coated samples were observed in SEM using secondary electron imaging at the accelerating voltage of 3 kV.

## 2.5 Data processing and statistical evaluation

The most of data processing, all figures, and all statistical calculations were made with a freeware Python programming language and its open-source data processing packages (NumPy, Matplotlib, SciPy, and Seaborn [46]). Some data pre-processing was made with a commercial spreadsheet program (MS Excel). IR spectra were processed with our own module MPINT, which can calculate all above-mentioned IR indexes (section 2.4.1) in an automated and reproducible way (see also section 3.3.2below). Within scatterplot matrix graphs (produced with Python seaborn module), we calculated and plotted the values of Pearson's correlation coefficients *r* and *p* values. Briefly, Pearson correlation coefficient (frequently denoted as Pearson's *r* or just *r*) is a measure of the linear correlation between two variables. Pearson's *r* ranges from +1 to -1, where +1 is a perfect positive linear correlation, -1 is a perfect negative linear correlation, and 0 means no linear correlation. Probability value (also referred to as *p*-value or just *p*) gives the probability for a given statistical model (here: correlation between the variables) that we would obtain the same or stronger result just by coincidence (here: that we would obtain the same or stronger correlation if the data were random). The results are considered statistically significant, if the *p*-value is below a significance level α, where α is usually 5 %, which means that the *p*-value of a statistically significant result should be < 0.05. More details can be found in the textbooks of statistics [47].

# 3 Results

## 3.1 Structure and properties of new UHMWPE liners

### 3.1.1 Oxidative degradation

Figure 1 summarizes the key UHMWPE properties related to an immediate oxidation (oxidation index from IR microspectroscopy; Fig. 1a, Eq. 1) and a long-term oxidation resistance (oxidation onset temperature from TGA analysis; Fig. 1b, section 2.4.3). Red columns denote the samples containing residual radicals, either from an irradiation without remelting or from sterilization with ionizing radiation. Orange columns represent samples, which contain *both* residual radicals *and* a stabilizer (usually α-tocopherol, which is the most active component of vitamin E); the only exception is sample K11 that contains an alternative commercial biocompatible stabilizer named Covernox$^{TM}$ (Table 1). Green column marks the only sample containing *no* residual radicals *and* a stabilizer. Grey samples comprise all other UHMWPE types, i.e. those without residual radicals and stabilizer.

Figure 1

The results in Fig. 1a confirmed that residual radicals increase the immediate oxidative degradation regardless of the stabilizer presence (the highest OI values for both red and orange columns). The common samples (grey columns) exhibited slightly lower OI values. The lowest OI value was observed for the stabilized sample without residual radicals (green sample). However, is should be noted that the oxidative degradation in all new samples was quite low (all OI values were well below 0.2, while the strong UHMWPE oxidation starts for OI values > 1 and critical oxidation for OI > 3 according to review of Kurtz et al. [48].

The results in Fig. 1b documented that the presence of a stabilizer increases the long-term oxidative resistance (the highest OOT temperatures for green and orange samples). All other samples exhibited lower OOT temperatures, i.e. their started to degrade at lower temperature than the stabilized sample. It is worth noting that TGA experiments are just a rough estimate of long-term oxidative resistance and more precise information is obtained from the sample behavior after the accelerated aging, as documented in section 3.2.

The presence of residual radicals in all samples marked by red and orange columns was confirmed by ESR spectroscopy. At the same time, the ESR spectroscopy documented that the samples marked by green and grey columns contained no detectable concentration of radicals. The ESR results are summarized in section 2 of the Supplementary information file. The complete numerical results of all experiments, including ESR measurements, can be found in the second Supplementary information file, which comes in the form of a large MS Excel table.

### 3.1.2 Degree of crosslinking

Figure 2 shows the properties that are connected with the degree of UHMWPE crosslinking: the trans-vinylene index from IR microspectroscopy (VI in Eq. 2; Fig. 2a) and the crosslinking density ($N_e$ in Eq. 15;

Fig. 2b). The values of VI are proportional to the total absorbed radiation dose for given experimental conditions and to concentration of crosslinks on condition that the irradiation is performed at suitable conditions [33]. The values of $N_e$ tend to increase if crosslinked UHMWPE molecules form insoluble polymer network [49].

Figure 2

The columns in Fig. 2 are marked according to the total absorbed radiation dose. The highly crosslinked samples (total dose ≥ 100 kGy) are marked with dark blue, the moderately crosslinked samples (total dose ≥ 50 kGy but lower than 100 kGy) are marked with medium blue, and the weakly crosslinked samples (just sterilized with ionizing radiation, where typical doses are around 30 kGy) are marked with light blue. All other, non-crosslinked samples are marked grey. The results confirmed that the total radiation dose correlates with both VI (Fig. 2a) and $N_e$ (Fig. 2b), although there were some variations that could be attributed to different crosslinking conditions and/or subsequent modifications applied by different manufacturers.

### 3.1.3 Crystallinity and micromechanical properties

Figure 3 displays the results of the crystallinity measurements (Figs. 3a–b) and microhardness measurements (Figs. 3c–d). The crystallinity and microhardness were obtained from four different methods: the crystallinity from IR (Eq. 3) and DSC (Eq. 4) and the microhardness from MH (Eq. 5) and MHI (Eq. 7). Numerous studies [7,13,31] have demonstrated that UHMWPE crystallinity is influenced by thermal treatment: it increases after annealing (heating slightly below the melting temperature, $T_m$) and decreases after remelting (heating slightly above the $T_m$). In Fig. 3, the annealed samples are marked with blue color, the remelted samples are marked with red color, and the samples without thermal treatment are grey. Furthermore, the pioneering studies of Balta-Calleja and co-workers [50,51] have shown that microhardness of semicrystalline polymers, including UHMWPE, is proportional to their overall crystallinity.

Figure 3

The results in Fig. 3 are in reasonable agreement with the general trends described in the previous paragraph. At first, the direct proportionality between crystallinity and microhardness is clearly visible from the visual comparison of Figs. 3a–b and Figs. 3c–d. At second, the variations in the crystallinity values from IR (Fig. 3a) and DSC (Fig. 3b) were quite small. The notable exception is sample K17, whose IR-estimated crystallinity (IR/CI) was lower than expected for an annealed sample, lower than the DSC-calculated crystallinity (DSC/CR), and lower than we could assume from the relatively high microhardness values (MH/$H_V$ and MHI/$H_{IT}$). Nevertheless, IR profile of sample K17 was re-measured twice (i.e. after the original measurement we performed additional two measurements from another location) and the surprisingly low IR/CI value was confirmed. This suggests that the estimation of crystallinity from IR spectroscopy may not be 100% reliable for specific cases. Finally, the average crystallinity of annealed samples (blue columns) tended to be higher than the average crystallinity of remelted samples (red columns), although the difference was not too high. Evidently, the individual variations among the different samples played quite important role and thermal treatment was just one of the factors influencing the final crystallinity values.

## 3.2 Structure and property changes in aged UHMWPE liners

### 3.2.1 Oxidative degradation

Figure 4 shows the oxidative degradation of the UHMWPE liners after the accelerated aging. The column colors are the same as in Fig. 1, which shows the oxidation of the samples before the aging. However, Fig. 1 gives just one value of oxidation index (overall oxidation index IR/OI, as the samples before aging were homogeneous), while Fig. 4 displays two OI values: (i) IR/OI(max) represents the maximal measured value close to the surface and (ii) IR/OI(ave) represents the average value in the middle of the aged specimen.

Figure 4

The oxidative damage of the samples before the accelerated aging (Fig. 1) and after the accelerated aging (Fig. 4) are rather different. At first, the OI values in Fig. 1 are quite low, which confirms that none of the modern UHMWPE liners was completely bad – at least at the beginning of its lifetime. At second, the OI values of the new samples (Fig. 1) correlate mostly with the presence of residual radicals (the high OI values for samples with radicals = orange and red columns), while the OI values of the aged samples (Fig. 4) correlate mostly with the stabilization (the low OI values for stabilized samples = green and orange columns). At third, the oxidative degradation close to the surface (Fig. 4a) was considerably stronger than the degradation inside the sample (Fig. 4b). Finally, the oxidative degradation on the surface of the aged samples was less modification-sensitive (Fig. 4a, lower difference between red and grey columns), while the degradation inside the specimen was strongly influenced by the presence of residual radicals and stabilizer (Fig. 4b, very high difference between red and grey columns). This suggests that the surface and bulk oxidation may have different mechanism.

Figure 5

Figure 5 displays the complete IR data for two selected UHMWPE samples after accelerated aging and illustrates the principle of data processing. Moreover, the figure documents the clear spectral difference between modern, well-stabilized and almost non-oxidized UHMWPE formulation (Fig. 5a–b; sample K02) and older UHMWPE formulation, which suffers from heavy oxidative degradation (Fig. 5c–d; sample K07). The data are shown as 2D-IR images (heatmaps in the center of each subfigure), from which we can extract individual 1D-IR spectra (upper subplots of each subfigure) or normalized absorbances of selected IR bands (right subplots of each subfigure). In fact, the right subplots with the normalized absorbances are the IR profiles (section 2.4.1) that are calculated automatically by our MPINT program (section 3.3.2). The main difference between the modern UHMWPE formulation (Figs. 5a–b) and older UHMWPE formulation (Fig. 5c–d) is the strong oxidation band around 1720 cm$^{-1}$. This band (C=O stretching vibration) represents the main UHMWPE oxidative degradation products, which are carbonyl compounds (namely ketones, esters, and carboxylic acids). Additional differences can be observed in the region 900–1000 cm$^{-1}$ (Figs. 5a and 5c) where we can see C=C bands (related to crosslinking as discussed in section 3.2.2) together with a band at 940 cm$^{-1}$ (Fig. 5c), which is typical of strongly oxidized samples. The 940 cm$^{-1}$ band was observed for all strongly oxidized UHMWPE formulations aged *in vitro* within this contribution, but also for the *in vivo* aged UHMWPE liners from our recent study [30]. Moreover, the careful comparison of the above-mentioned *in vitro* and *in vivo* aged UHMWPE samples revealed that their IR spectra exhibit, in addition to the dominating C=O band around 1720 cm$^{-1}$, three additional weaker bands: (i) the 940 cm$^{-1}$ band discussed above, (ii) a small 1405 cm$^{-1}$ shoulder, and (iii) a weak 3530 cm$^{-1}$ peak (the typical IR spectra with all four oxidation-related peaks are shown in section 1 of the Supplementary information file).

### 3.2.2 Degree of crosslinking

Figure 6 documents that the accelerated aging did not influence the values of VI, which are proportional to the degree of crosslinking. Nevertheless, the crosslinking density after the accelerated aging may decrease as a result of oxidative degradation, which leads to polymer chain scissions. This is well documented in the literature [11,52] and briefly discussed below (section 4.3).

Figure 6

Figure 7 shows the IR spectra of four typical samples: (i) *non-crosslinked and stabilized* (sample K15; Fig. 7a), (ii) *non-crosslinked and non-stabilized* (sample K09; Fig. 7b), (iii) *crosslinked and stabilized* (sample K04; Fig. 7c), and (iv) *crosslinked and non-stabilized* (sample K17; Fig. 7d). The term stabilized sample means that the residual radicals from the ionizing radiation are not present or they have been eliminated. This could be due to one of the following reasons: (a) the sample was not modified by ionizing radiation at all, (b) the sample was modified by the ionizing radiation, but the residual radicals were removed completely by remelting or (c) the residual radicals were suppressed by a biocompatible stabilizer such as vitamin E. The detailed description of all samples is in Table 1. Each of the 21 samples in Table 1 falls into one of the four abovementioned categories, which are represented by four typical IR spectra in Fig. 7.

Figure 7

From the point of view of oxidation and crosslinking, the IR spectra of UHMWPE samples comprise two important regions: (i) region from 1600 to 1800 cm$^{-1}$ with oxidation-related bands corresponding to C=O bonds and (ii) region from 900 to 1000 cm$^{-1}$ with bands corresponding to C=C bonds. The non-crosslinked and stabilized sample (Fig. 7a) may show a very small peak at 910 cm$^{-1}$ (terminal vinylene groups, which may be present in non-crosslinked UHMWPE) and no strong oxidation peaks around 1720 cm$^{-1}$ (low to negligible concentration of carbonyl groups, which are the main oxidation product in UHMWPE) [53]. The non-crosslinked and non-stabilized sample (Fig. 7b) shows two small peaks at 940 cm$^{-1}$ and 965 cm$^{-1}$ (the former is typical of strongly oxidized samples and the latter is attributed to in-chain trans-vinylene group that is formed as a by-product when UHMWPE is treated with ionizing radiation [33,54]) and one strong oxidation peak around 1720 cm$^{-1}$ (resulting from oxidative degradation due to residual radicals [49,55]). The crosslinked and stabilized sample (Fig. 7c) shows a peak at 965 cm$^{-1}$ (in-chain trans-vinylene group that is formed as a by-product during radiation-induced crosslinking [10,56]), no other peaks in 900–1000 cm$^{-1}$ region (terminal vinylene groups with peak at 910 cm$^{-1}$ are consumed during crosslinking and 940 cm$^{-1}$ peak is not present because stabilization prevents strong oxidation) and no strong peak in 1600–1800 cm$^{-1}$ region (just a minimal oxidation in well-stabilized samples without residual radicals). The crosslinked and non-stabilized sample shows two peaks at 940 cm$^{-1}$ and 965 cm$^{-1}$ (in analogy with the non-crosslinked and non-stabilized sample in Fig. 7b, the former peak indicates strong oxidation and the latter the interaction of the sample with ionizing radiation during crosslinking and/or sterilization) and a strong peak around 1720 cm$^{-1}$ (in analogy with the non-crosslinked and non-stabilized sample in Fig. 7b, the residual radicals cause strong oxidative degradation).

### 3.2.3 Crystallinity and micromechanical properties

Figure 8 combines the results of the crystallinity measurement from IR (Fig. 8a) and instrumented microindentation (Figs. 8b–d). The colors are the same as in Fig. 4, because the decisive parameter influencing crystallinity and micromechanical properties of the aged samples is the oxidative degradation. All non-stabilized samples with residual radicals (red columns) suffered from the highest oxidative degradation (as documented in Fig. 4), which resulted in the highest crystallinities (as proved in Fig. 8a). The highest crystallinities correlated with the highest values of hardness (Figs. 8b and 8c) and elastic modulus (Fig. 8d). The observed *oxidation-crystallinity-hardness-modulus* correlation were in perfect agreement with theoretical predictions [50] and our previous studies [30,39]. The stabilized samples with residual radicals (Fig. 8; orange columns) and the stabilized sample without residual radicals (Fig. 8; green column) were not too different from all other samples (Fig. 8; gray columns). This confirmed that a sufficient concentration of a biocompatible stabilizer can eliminate the impact of residual radicals on oxidative degradation (Fig. 4) and consequently on the local mechanical properties (Fig. 8) of UHMWPE.

Figure 8

Figure 9 further illustrates the close relations among the sample modifications, oxidation, morphology, and mechanical properties. Typical modern UHMWPE's (represented by sample K16 in Fig. 9a) are crosslinked and stabilized by remelting and/or biocompatible stabilizers. The older types of UHMWPE (represented by sample K19 in Fig. 9b) contain residual radicals from crosslinking or sterilization and no biocompatible stabilizer.

Figure 9

The difference between the cut surfaces of the two samples in Fig. 9 is clearly visible. The cut surface of the modern, stabilized UHMWPE is smooth, with a uniform morphology and no signs of the oxidative degradation at the edges. The imprints of the indenter on the surface are bigger, which indicates that the sample is softer (i.e. no hardening due to oxidation-induced cold crystallization [30]). The cut surface of the older, non-stabilized UHMWPE with residual radicals is less smooth, and clearly degraded on the surface. The edges are evidently more brittle as they show rough morphology (brittle fragments released from the surface) and chatter (unwanted vibrations during the cutting, typical of brittle polymer materials). The imprints of the indenter on the cut surface of the non-stabilized sample are smaller, which could be attributed to oxidation induced chain scissions that lead to cold crystallization followed by an increase in the overall crystallinity stiffness, hardness and brittleness [30,49]. The rough surface morphology of the heavily oxidized samples was the reason why the micromechanical properties could not be measured directly from the surface of the aged samples; the morphology and properties of UHMWPE surfaces after the accelerated aging are given in section 3 of the Supplementary information file.

## 3.3 Additional results

### 3.3.1 New method of UHMWPE aging

The reliable accelerated aging method is inevitable to predict long-term properties of UHMWPE. In this work, we introduced our own method, simulating the *in vivo* UHMWPE aging. The new method, whose technical details of are described above in section 2.3, is a moderate modification of previously described protocols employing the $H_2O_2$ accelerated aging of UHMWPE [48,57], but it exhibits two important advantages: (i) It is faster as it requires 30 days of aging while the previous solution-based protocols require >60 days to achieve comparable oxidation levels [57,58]. (ii) In the IR spectra, our method shows no *non-physiological oxidation peaks* attributed to aldehydes at 1732 $cm^{-1}$, which may appear during less bio-compatible accelerated aging protocols [58,59]; the detailed comparison of IR spectra for a typical UHMWPE sample after our *in vitro* aging with UHMWPE sample retrieved after 11 years of *in vivo* aging is given in section 1 of the Supplementary information file.

The limitation of our accelerated aging method consists in that the $H_2O_2$ aged samples exhibit the oxidation maximum at the surface, while the *in vivo* and shelf-aged UHMWPE samples show the oxidation maximum in the subsurface region [16,34]. Nevertheless, the two most widely used methods of UHMWPE accelerated aging, which are oxygen bomb (2 weeks at 70 °C under 5 bar oxygen) and convection oven (3 weeks at 80 °C in the air circulating oven) [60], suffer from similar problems [53] and yield lower levels of oxidation [57]. The reason why the available methods fail to produce the subsurface oxidation peak seems to result from the fundamental differences in oxidative mechanism between *in vivo* and laboratory conditions. In vivo, the oxygen diffuses and reacts with the polymer over long time, while the accelerated aging protocols expose the material to higher oxygen levels in much shorter time, leading to surface oxidation rather than to a distinct subsurface oxidation peak. To get more realistic oxidation profiles, it is necessary to employ more sophisticated and longer protocols reflecting in vivo conditions (synovial fluid, cyclic loading, and longer times for oxygen diffusion) or to use real-time aging [61,62].

### 3.3.2 MPINT: Updated software for automated evaluation of IR spectra

IR spectroscopy is a key method for the evaluation oxidation and structure changes in UHMWPE liners. In a typical experiment, the liner is cut perpendicular to the exposed surface and IR spectra are measured in the form of line scan as described in the Experimental section above. The line scan is usually measured with a step of 100 µm, which represents tens of spectra per sample. As each sample is measured at least twice to get more reproducible results and we have >20 samples in this study, the whole dataset comprises hundreds of IR spectra. From each spectrum, we need to evaluate the values of indexes (OI, CI, VI, and EI; section 2.4.1). In order to make the spectral processing easier, faster, and more reproducible, we developed MPINT software (Multiple Peak INTegration; section 2.5; ref. [30]). It is an open-source Python package, which can be added to arbitrary Python installation using the standard www-repository known as PyPI (Python Package Index; the website of our package is *https://pypi.org/project/mpint*). Within this work, the package was updated and its documentation was improved. All IR spectra in this contribution were auto-processed by MPINT, which should make the final values less user-dependent and thus more reproducible. The details about the package can be obtained in the abovementioned website.

### 3.3.3 Complete set of micromechanical properties including creep and elasticity

All 21 UHMWPE samples in this contribution were characterized by a microindentation hardness testing in its non-instrumented (MH) and instrumented (MHI) form. The micromechanical properties can be divided into two groups [41,63]: *stiffness-related properties* (elastic modulus, $E_{IT}$, and various types of hardness, $H_{IT}$, $H_M$, and $H_V$; Eq. 5–8) and *viscosity-related properties* (indentation creep, $C_{IT}$, and elastic part of the indentation work, $\eta_{IT}$; Eq. 9–10). For UHMWPE samples, the most relevant were the stiffness-related properties, which were influenced by both UHMWPE modifications (section 3.1.3) and oxidation-induced crystallinity changes (section 3.2.3). The main trend consisted in that the oxidative degradation increased the crystallinity and all stiffness-related properties ($E_{IT}$, $H_{IT}$, $H_M$ and $H_V$), which was accompanied by a small increase in creep ($C_{IT}$, the first viscosity-related property) and a small decrease in elasticity ($\eta_{IT}$, the second viscosity-related property). Briefly, the changes of the viscosity-related properties documented that the increase in crystallinity lead to higher plasticity (i.e. the crystalline domains were more difficult to deform, but the deformation was more permanent). The correlations between oxidation, crystallinity, and all micromechanical properties from MH and MHI measurements are given in section 4 of the Supplementary information file. The complete results of all measurements are given in the second Supplementary information file in the form of MS Excel table.

# 4 Discussion

## 4.1 Variability of contemporary UHMWPE formulations for TKR

The results of spectroscopic, thermal, micromechanical and microscopic methods, which were presented above in section 3, have demonstrated that the contemporary UHMWPE formulations employed in TKR exhibit surprisingly wide range of properties. In this section, the key differences among the UHMWPE formulations will be discussed in more detail. We will summarize the most important differences both before (Fig. 10) and after (Fig. 11) the accelerated aging.

Figure 10

Figure 10 documents that the properties of the original, non-aged UHMWPE formulations are influenced by multiple factors, such as stabilization (Fig. 10a), crosslinking (Fig. 10b) and thermal treatment (Fig. 10c). From the point of view of stabilization, the UHMWPE formulations can be divided into three groups: (i) samples with biocompatible stabilizers, (ii) samples without stabilizers and without residual radicals, and (iii) samples without stabilizers and with residual radicals. Most of the samples with biocompatible stabilizers (Fig. 10a, yellow box) contained residual radicals (samples K04, K08, K11, K16, K18, and K21; yellow bars in Fig. 1) and just one of them was without residual radicals (sample K02; green bar in Fig. 1). However, all samples with stabilizers (Fig. 10a, yellow box) were more resistant to oxidative degradation than samples without stabilizers and residual radicals (Fig. 10b, gray box) and samples without stabilizers (Fig. 10c, red box). The differences among all three groups were statistically significant ($p < 0.001$). As for the level of crosslinking (Fig. 10b), the highly crosslinked samples (Fig. 10b, dark blue box) exhibited higher values of trans-vinylene index than the moderately crosslinked samples (Fig. 10b, light-blue box) and non-crosslinked samples (Fig. 10c, gray box). The differences among all three groups were statistically significant ($p < 0.002$). The observed increase in VI as a function of the crosslinking density agreed with the literature [33,54]. Thermal treatment influenced mechanical performance of the samples (Fig. 10c), although the effect was weaker than it might have been expected. Most of the literature concludes that thermal treatment of UHMWPE influences its crystallinity and stiffness, i.e. elastic moduli, yield stress, and indentation hardness at all length scales [34,39]. The annealing and remelting of *one specific UHMWPE formulation* leads to the stiffness increase and decrease, respectively [31,64]. However, in our case we compared *21 different UHMWPE formulations*, whose initial stiffness depended on multiple processing and/or modification parameters. Consequently, the samples without thermal treatment exhibited intermediate stiffness as expected (Fig. 10c, gray box), but the average stiffness of the annealed samples was just slightly higher (Fig. 10c, orange box; the statistically insignificant difference with $p > 0.5$), and the average stiffness of the remelted samples just moderately lower (Fig. 10c, blue box; the significant difference with $p < 0.02$).

Figure 11

Figure 11 illustrates that the key properties of the aged UHMWPE samples depended mostly on the stabilization, while the other parameters played a minor role. The samples with a biocompatible stabilizer (Fig. 11, yellow boxes) and the samples without residual radicals (Fig. 11, gray boxes) showed the lowest oxidative degradation (Fig. 11a), the lowest crystallinity (Fig. 11b), and the lowest hardness (Fig. 11c). The samples with residual radicals (Fig. 11, red boxes) exhibited much higher oxidative degradation (Fig. 11a), which resulted in higher crystallinity due to cold crystallization (Fig. 11b), and higher hardness (Fig. 11c). Therefore, the stabilization and/or absence of the residual radicals prevented the long-term changes of the UHMWPE structure and properties, while the presence of residual radicals led to the severe oxidative degradation and the corresponding changes of the material.

## 4.2 Correlation between structure and mechanical properties of UHMWPE in TKR

UHMWPE is a semicrystalline polymer, whose stiffness-related properties (such as elastic modulus, yield strength, and hardness) depend mostly on the overall crystallinity ($v_c$, volume fraction of crystalline phase) [50,51]. In fact, the linear correlation between the crystallinity and stiffness-related properties of semicrystalline polymers is such a reproducible trend that it can be employed as a criterion that verifies the reliability of the characterization methods used [39]. The crystallinity-stiffness relation can be summarized in the form of Eq. 14:

$$E \approx E_{IT} \approx 30Y \approx 10H_M \approx 10H_{IT} \propto v_c \tag{14}$$

where $E$ is the macroscale elastic modulus, $E_{IT}$ is the indentation modulus, $Y$ is the macroscale yield strength, $H_M$ is the Martens hardness (also known as universal hardness), $H_{IT}$ is the indentation hardness, and $v_c$ is the volume fraction of crystalline phase. The seemingly simple Eq. 14 is based on extensive theory, which was elaborated in the pioneering studies of Tabor [65], Balta-Calleja and co-workers [66], Oliver and Phar [36], and Struik [67]. Simplified justification of Eq. 14 with the relevant references can be found in our recent contributions [39,63]. More details can be found in book on polymer micromechanics [41,68]. In the case of UHMWPE liners, the additional important parameter is the oxidative degradation, which increases the crystallinity due to chain scissions followed by cold crystallization, as exemplified above (section 3.2.3) and discussed in detail elsewhere [30,49]. Within this contribution, we estimated the oxidation and crystallinity using IR microspectroscopy (as described in section 2.4.1) and measured the micromechanical properties by MHI testing (as described in section 2.4.4). The final structure-properties correlations are shown in Fig. 12.

Figure 12

Figure 12 summarizes the theoretically predicted correlations between oxidative degradation (IR/OI), crystallinity (IR/CI), and two stiffness-related micromechanical properties (MHI/$E_{IT}$ and MHI/$H_{IT}$) for all UHMWPE samples. The oxidation-crystallinity-stiffness proportionality is evident. Moreover, the $p$ values (shown in the lower right corner of each subplot in Fig. 12) prove that all correlations are statistically significant. This confirms the reliability of our microscale characterization methods, which yielded the consistent results in line with the theory represented by Eq. 14. According to Pearson's coefficients $r$ (shown in the upper left corner of each subplot in Fig. 12), the correlations before the aging (Fig. 12a) are somewhat weaker than the correlations after the aging (Fig. 12b). This can be attributed to two facts: (i) The crystallinities of all UHMWPE samples before the aging are quite similar to each other and so the trends are not so clear. (ii) The crystallinities of the UHMWPE samples before the aging are influenced by oxidation as well as other factors, such as thermal treatment, while the crystallinities of the UHMWPE samples after the aging are influenced mostly by oxidative degradation, while the other factors pay a minor role (as already discussed in section 4.1).

## 4.3 Influence of UHMWPE modifications on properties before and after aging

In the results section, it has been demonstrated that the contemporary UHMWPE formulations for TKR represent a broad range of materials with different chemical, physical and mechanical properties (section 3.1; Figs. 1–3). Moreover, the differences among the various UHMWPE types tended to amplify after the accelerated aging (section 3.2; Figs. 4–9). The collected data enabled us to generalize how the key UHMWPE modification parameters (irradiation, thermal treatment, stabilization, and sterilization) influenced the polymer properties before and after the accelerated aging:

- *Irradiation* of UHMWPE with gamma rays or accelerated electrons under suitable conditions (such as inert atmosphere, high dose rates etc.) results in crosslinking, which increases the wear resistance [52,54]. The degree of crosslinking is proportional to the total absorbed radiation dose (from both irradiation and sterilization). It can be estimated from solubility measurements (crosslinking density, $N_e$) and IR spectroscopy (trans-vinylene index, VI) [33,49]. Before the accelerated aging, all UHMWPE samples showed a reasonable correlation among total absorbed dose, *g* and VI (Fig. 2 and 10b); certain variations could be attributed to different conditions during the irradiation of different UHMWPE formulations. After the accelerated aging, the VI values remained almost the same (Fig. 6), but this just indicated that the concentration of C=C bonds (a side product of irradiation) was unchanged, while the degree of crosslinking may have changed. Indeed, the non-stabilized samples with residual radicals suffered from heavy oxidative degradation (Fig. 4, red columns), which caused chain scissions, cold crystallization, and unwanted increase in stiffness-related properties (Fig. 8, red columns); these changes are known to be connected with a decrease in the crosslinking density and wear resistance [11,52].
- *Thermal treatment* influences both molecular and supermolecular structure of UHMWPE formulations [13,52]. As for molecular structure, the most important is the elimination of residual radicals, which can be either none (in the case of no thermal treatment; NN), partial (in the case of annealing; AN), or complete (in the case of remelting; RM). Before the accelerated aging, the irradiated samples without the complete removal of radicals (NN and AN samples) and without a stabilizer (usually α-tocopherol) showed an increased immediate oxidative degradation (Fig. 1a, red columns) and a lower long-term oxidation stability (Fig. 1b, red columns). The irradiated samples with a stabilizer showed an increased immediate oxidative degradation as well (Fig. 1a, orange columns), but they exhibited a high long-term oxidation stability (Fig. 1b, orange columns). Therefore, the UHMWPE stabilization was more important than the thermal treatment from the point of view of long-term oxidation resistance. This was confirmed by the results for the samples after the accelerated aging, in which the *non-stabilized NN/AN-samples* with residual radicals exhibited by far the strongest oxidative degradation associated with a cold crystallization and stiffening (Figs. 4 and 8, red columns), while the *stabilized NN/AN-samples with residual radicals* exhibited much lower levels of oxidative degradation and subsequent structure changes (Figs. 4 and 8, orange columns). As for supermolecular structure, the thermal treatment influenced crystallinity and micromechanical properties of the samples before aging (Fig. 3), which agreed with literature [30], but the differences among the samples were small (Fig. 10c) and completely overridden by the oxidative degradation after the accelerated aging (Figs. 8, 9 and 11).
- *Stabilization* of UHMWPE increases its resistance to oxidation strongly. The stabilization did not influence structure and properties of the polymer before the aging too much (Fig. 10), but it showed to be the crucial factor for the aged samples (Fig. 11). Before the aging, the presence of the stabilizer impacted mostly on OOT temperature (Fig. 1, orange and green columns), but this parameter is linked with the long-term oxidation stability, i.e. with the possible material changes after aging rather than with the current structure and properties. After the aging, the stabilized samples exhibited the lowest oxidation close to the surface and inside the specimen (Fig. 4, green and orange columns). Consequently, the stabilized samples did not suffer from the structure and properties changes after accelerated aging (Fig. 8, compare the data of the stabilized samples marked green and orange with the data of non-stabilized samples marked red). Moreover, the difference between stabilized and non-stabilized UHMWPEs was clearly visible on the LM and SEM micrographs (Fig. 9). Finally, the differences between the stabilized samples and non-stabilized samples with residual radicals were statistically significant for molecular structure (OI, Fig. 11a), supermolecular structure (CI, Fig. 11b), and mechanical properties ($H_M$, Fig. 11c).

- *Sterilization* used to be a critical issue for historical UHMWPE formulations based on non-modified, gamma-sterilized polymers, in which the residual radicals from sterilization could cause oxidative degradation and early failures [69,70]. Modern sterilization methods, namely ethylene oxide and gas plasma sterilization, do not generate the residual radicals, which eliminated the problem [14,15]. Surprisingly enough, some manufactures continue selling the gamma-sterilized (or electron beam-sterilized) UHMWPE formulations without stabilization as of the year 2020 (Table 1, samples K05, K07, K09, K17, and K19); all these samples showed the highest oxidative degradation after the accelerated aging (Fig. 4, red columns) with the corresponding negative structure and properties changes (Fig. 8, red columns). On the other hand, if the gamma-sterilization is combined with a sufficient stabilization in modern 2nd generation crosslinked UHMWPEs (Table 1, samples K04, K08, K11, K18, and K21), the residual radicals are eliminated and the samples show very good oxidation stability (Fig. 4, orange columns).

# 5 Conclusion

We have compared the structure and properties of 21 contemporary UHMWPE liners used in modern total knee replacements. The contemporary UHMWPE formulations for TKR cover a broad range of materials, from traditional polymers with minimal modifications, through the 1st generation crosslinked UHMWPEs, to the most recent 2nd generation crosslinked UHMWPEs with biocompatible stabilizers. All samples were obtained from their manufactures, in the form of tibial components, whose limited size and irregular shape forced us to employ microscale characterization methods, the most important of which were IR microspectroscopy and microindentation. Each sample was characterized both before and after the accelerated aging in $H_2O_2$. The main conclusions can be summarized as follows:

1. The UHMWPEs used in contemporary TKR exhibit quite broad range of chemical, physical, and mechanical properties.
2. The properties of the UHMWPE samples before the accelerated aging depend on multiple modification parameters, namely the degree of crosslinking, type of thermal treatment, presence of a stabilizer, and type of sterilization.
3. The properties of the UHMWPE samples after the accelerated aging are influenced mostly by the oxidative degradation, while the other factors are of minor importance. This underlined the importance of biocompatible stabilizers – all stabilized UHMWPE formulations showed significantly higher resistance to the long-term oxidative degradation.

4. The samples were characterized in microscale using various spectroscopic, micromechanical, thermal, chemical, and microscopic methods. Numerous correlations among the results from the independent methods, such the *oxidation-crystallinity-stiffness* proportionality, confirmed the reliability and reproducibility of our measurements.
5. The residual radicals in UHMWPE for TKR should be eliminated completely either by remelting or by a biocompatible stabilizer; the non-stabilized UHMWPEs with residual radicals, even if they were removed partially by annealing, should be avoided due to their insufficient long-term oxidation resistance.
6. The results of the accelerated aging in this contribution suggest that the best UHMWPE formulation for TKR should be crosslinked to get high wear resistance, stabilized by a biocompatible stabilizer to achieve long-term oxidation resistance and sterilized by ethylene oxide or gas plasma to minimize the concentration of residual radicals.

## Acknowledgement

This research was supported by AZV CR, project NU21-06-00084. IS thanks to project OP JAK Amulet, No. CZ.02.01.01/00/22_008/0004558, of the Ministry of Education, Youth and Sports of the Czech Republic, which is co-funded by the European Union. MS would like to thank to Dr. Jan Pilar for the measurement of ESR spectra.

## Research data for this article

All data are available at request to the corresponding author.

# 6 Figure 1

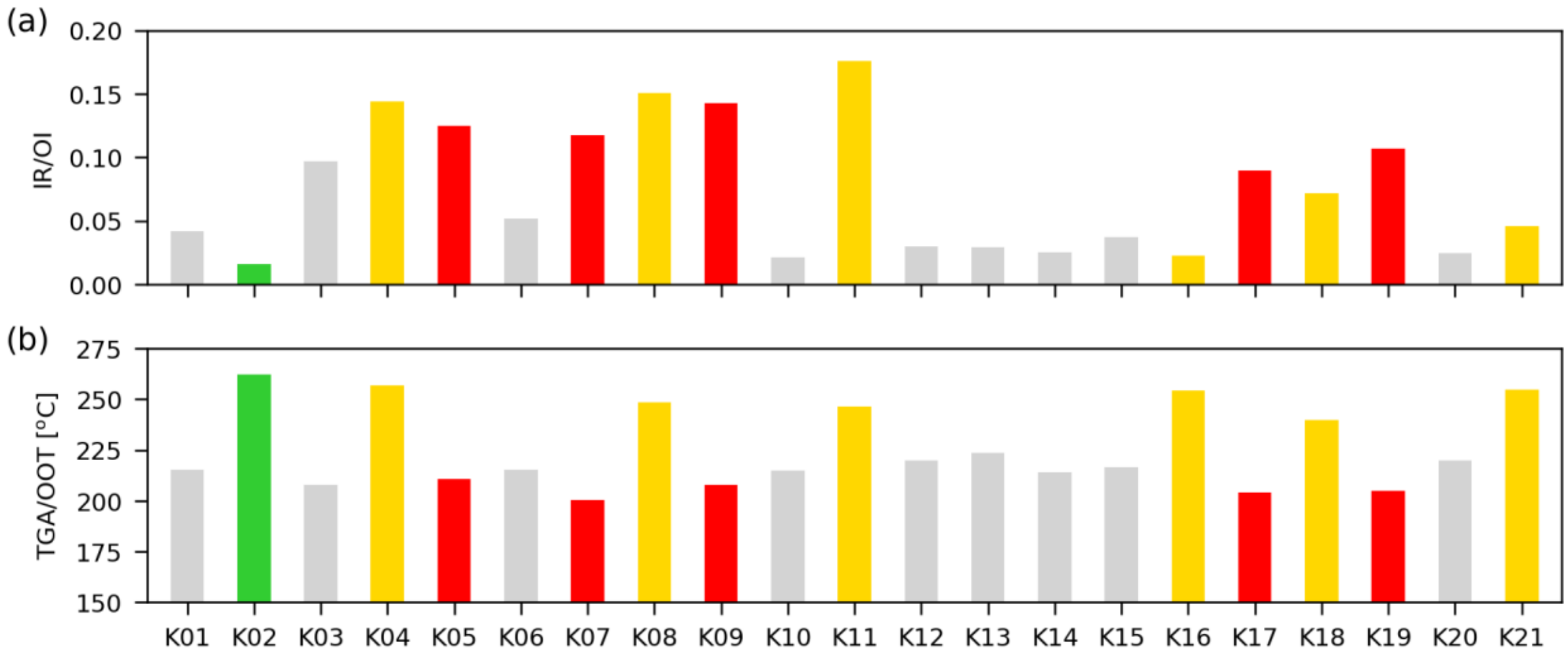

**Figure 1:** Oxidation index from IR (a) and oxidation onset temperature from TGA (b) for all UHMWPE formulations before accelerated aging. The column colors: red = samples with residual radicals and no stabilizer, orange = samples with residual radicals and a stabilizer, green = sample without residual radicals and with a stabilizer, gray = all other samples (without residual radicals and a stabilizer). Standard deviations: for IR/OI the standard deviations did not exceed 5% of the measured values; for TGA/OOT, each sample was measured just once (the second measurement only in the case of a suspicious result).

# 7 Figure 2

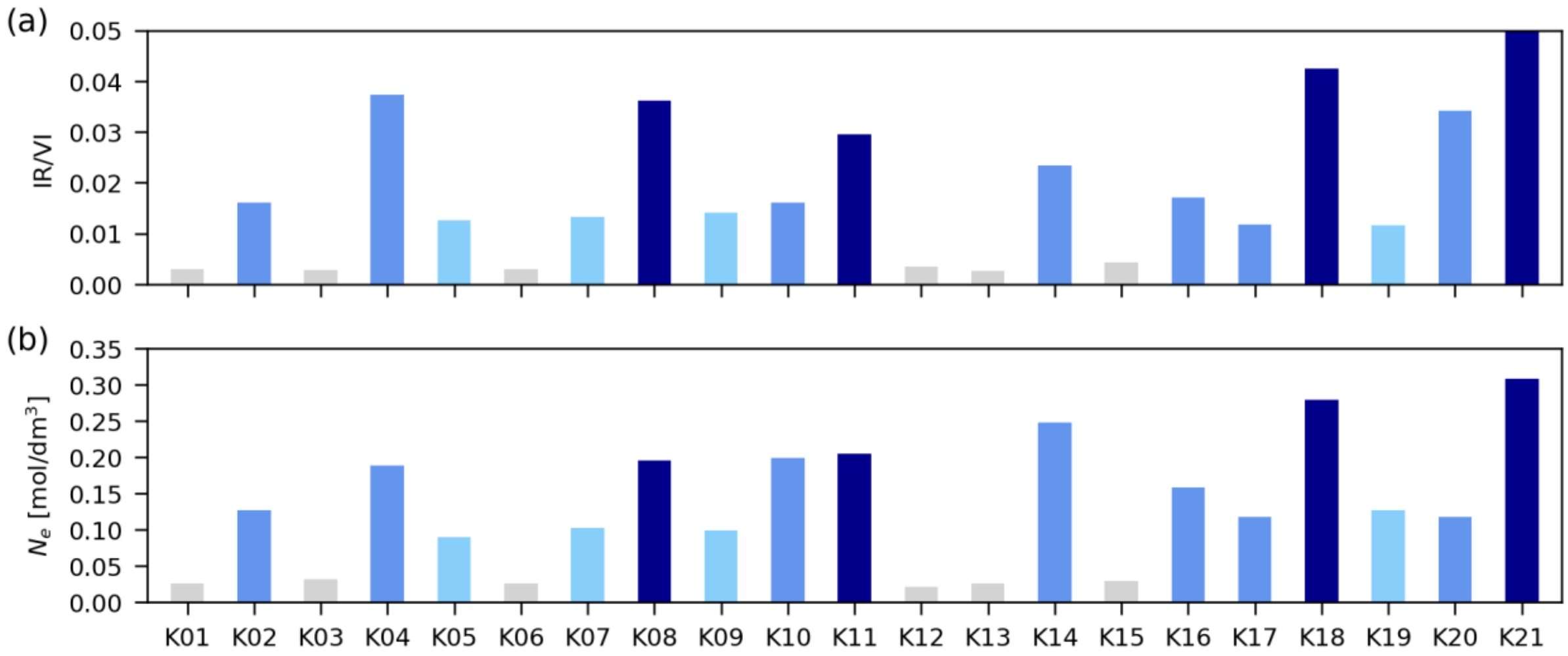

**Figure 2:** Trans-vinylene index from IR (a) and crosslinking density from swelling experiments (b) for all UHMWPE formulations before accelerated aging. The column colors: light blue = samples modified with low doses of ionizing radiation (total dose < 50 kG), medium blue = samples modified with medium doses of ionizing radiation (total dose between 50 and 100 kGy), dark blue = sample modified with high doses of ionizing radiation (total dose ≥ 100 kG), gray columns = all other samples (not modified with ionizing radiation). Technical note: for the low dose irradiated samples (light blue columns), the ionizing radiation was used just for sterilization, while for the medium and high dose irradiated samples (medium and dark blue columns) the ionizing radiation was used for crosslinking (and in some case for sterilization as well). Standard deviations: for IR/VI the standard deviations did not exceed 5% of the measured values; for solubility measurements, each sample was measured three times and the standard deviations were below 20% of the measured values.

# 8 Figure 3

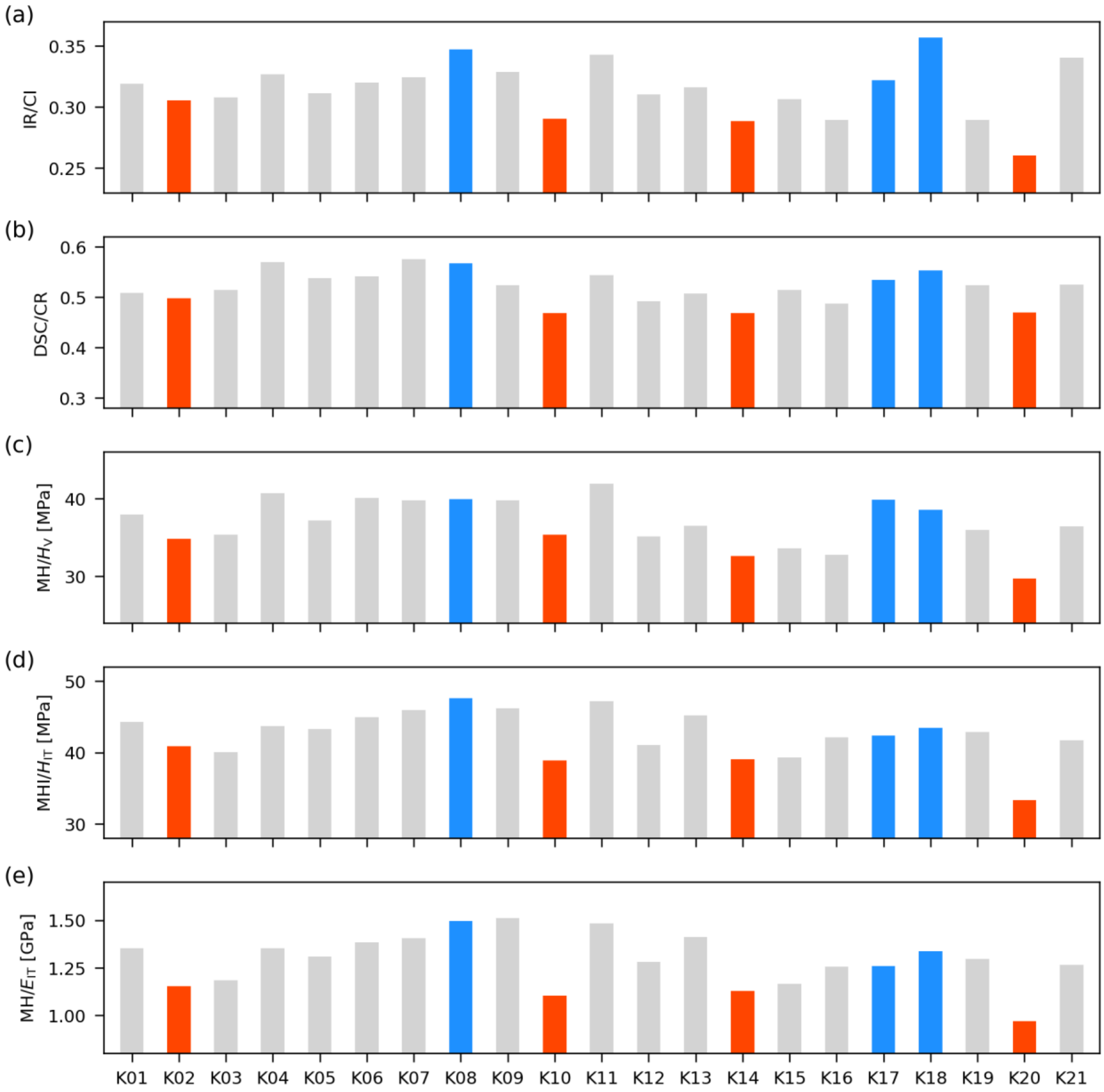

**Figure 3:** Crystallinity index from IR (a), crystallinity from DSC (b), Vickers hardness from non-instrumented microindentation (c), indentation hardness from instrumented microindentation (d), and indentation modulus from instrumented microindentation. The column colors: red = samples modified by remelting, blue = samples thermally modified by annealing, and gray = all other samples (without thermal treatment). The standard deviation did not exceed 5 % of the measured values.

# 9 Figure 4

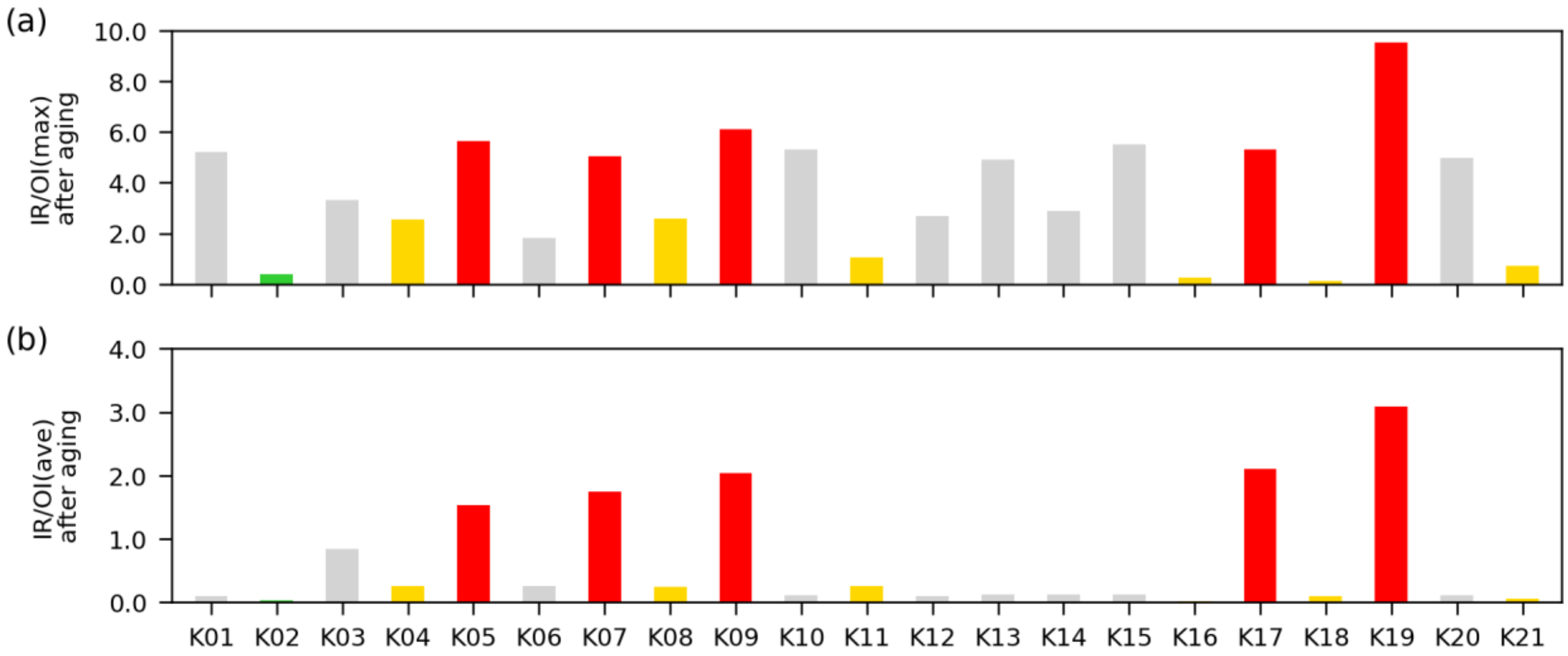

**Figure 4:** Maximal (a) and average (b) value of oxidation index from IR for all UHMWPE formulations after accelerated aging. The column colors are the same as in Fig. 1, which shows the samples before the accelerated aging: red = samples with residual radicals and no stabilizer, orange = samples with residual radicals and a stabilizer, green = sample without residual radicals and with a stabilizer, gray = all other samples (without residual radicals and a stabilizer). The standard deviations did not exceed 5% of the measured values.

# 10 Figure 5

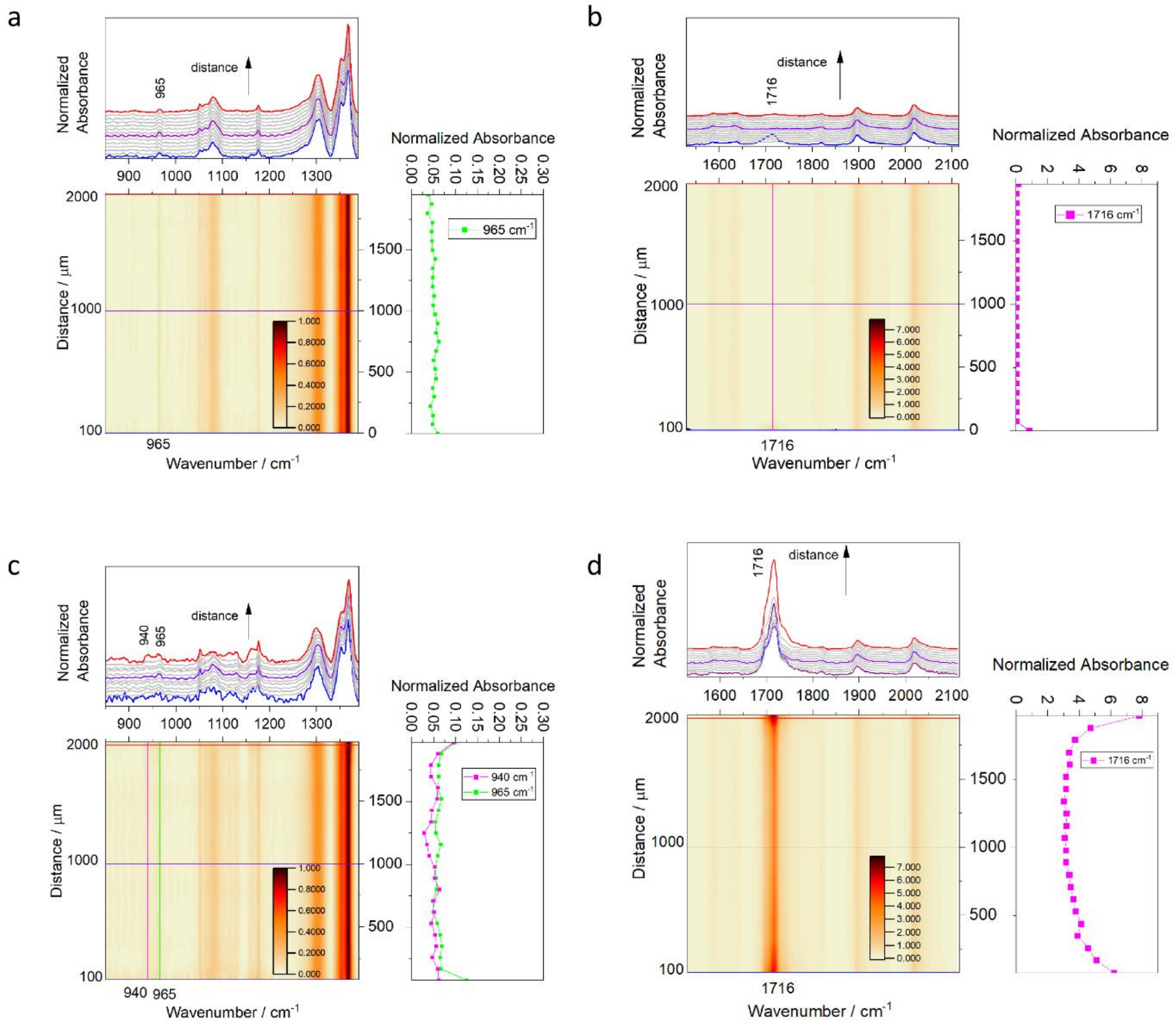

**Figure 5:** 2D images of IR spectra of two selected UHMWPE samples: (a, b) modern UHMWPE formulation without residual radicals (sample K02 from Table 1, which was crosslinked, remelted and stabilized with ethylenoxide) and (c, d) older UHMWPE formulation (sample K07 from Table 1, which was just sterilized by gamma-irradiation). Each of the four subfigures (a–d) consists of a 2D-heatmap showing IR intensities as a function of distance from the surface, top plot showing a series of representative IR spectra, and left plot showing the normalized absorbance of selected IR bands. All bands were normalized to the same peak at 1370 $cm^{-1}$. The wavelengths of key peaks are marked in each image.

# 11 Figure 6

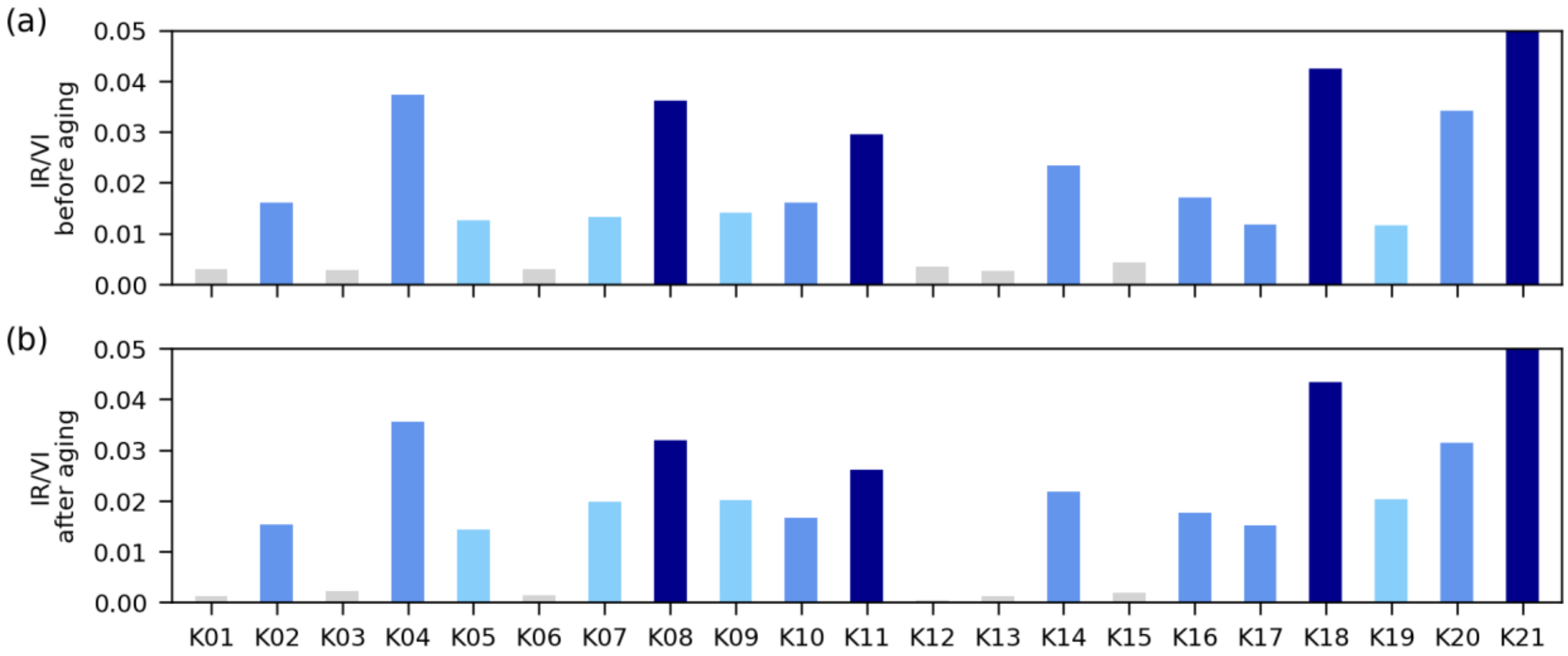

**Figure 6:** Trans-vinylene index from IR for all UHMWPE formulations before (a) and after (b) the accelerated aging. The column colors are the same as in Fig. 2: light blue = samples modified with low doses of ionizing radiation (total dose < 50 kG), medium blue = samples modified with medium doses of ionizing radiation (total dose between 50 and 100 kGy), dark blue = sample modified with high doses of ionizing radiation (total dose ≥ 100 kG), gray columns = all other samples (not modified with ionizing radiation). Technical note: for the low dose irradiated samples (light blue columns), the ionizing radiation was used just for sterilization, while for the medium and high dose irradiated samples (medium and dark blue columns) the ionizing radiation was used for crosslinking (and in some case for sterilization as well). The standard deviation of IR/VI measurements did not exceed 5% of the measured values.

# 12 Figure 7

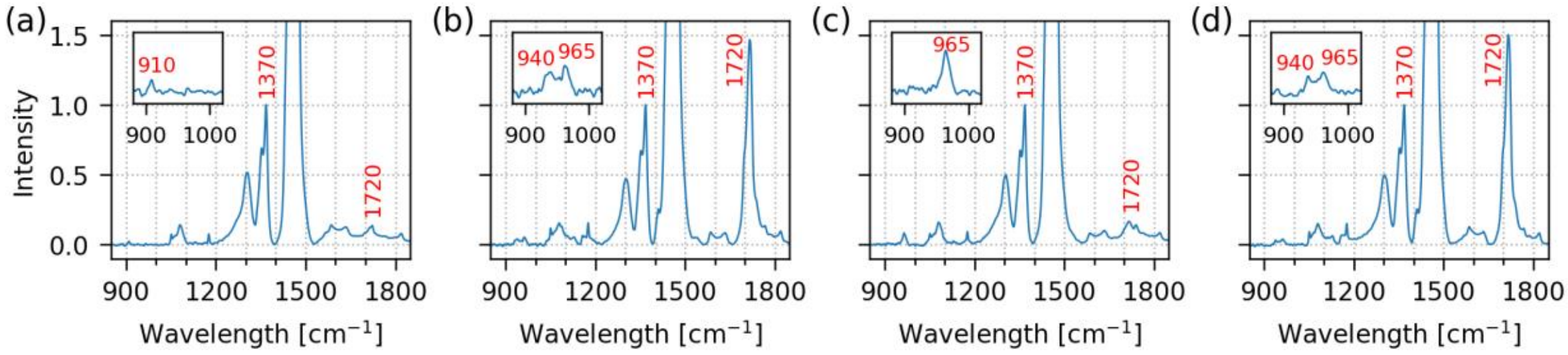

**Figure 7**: IR spectra of four typical UHMWPE samples: (a) sample K15 – non-crosslinked and without residual radicals, (b) sample K09 – non-crosslinked and with residual radicals, (c) sample K04 – crosslinked and stabilized so that the residual radicals were eliminated, and (d) sample K17 – crosslinked and not fully stabilized so that it contains residual radicals. More details about each sample can be found in Table 1. All spectra were normalized to the standard band at 1370 cm$^{-1}$. The main oxidation-related carbonyl band appears at 1720 cm$^{-1}$. More details about the typical UHMWPE peaks can be found in section 2.4.1 and references therein. The inset in each spectrum shows the region from 900–1000 cm$^{-1}$ with terminal vinylene band at 910 cm$^{-1}$ and in-chain trans-vinylene bad at 965 cm$^{-1}$.

# 13 Figure 8

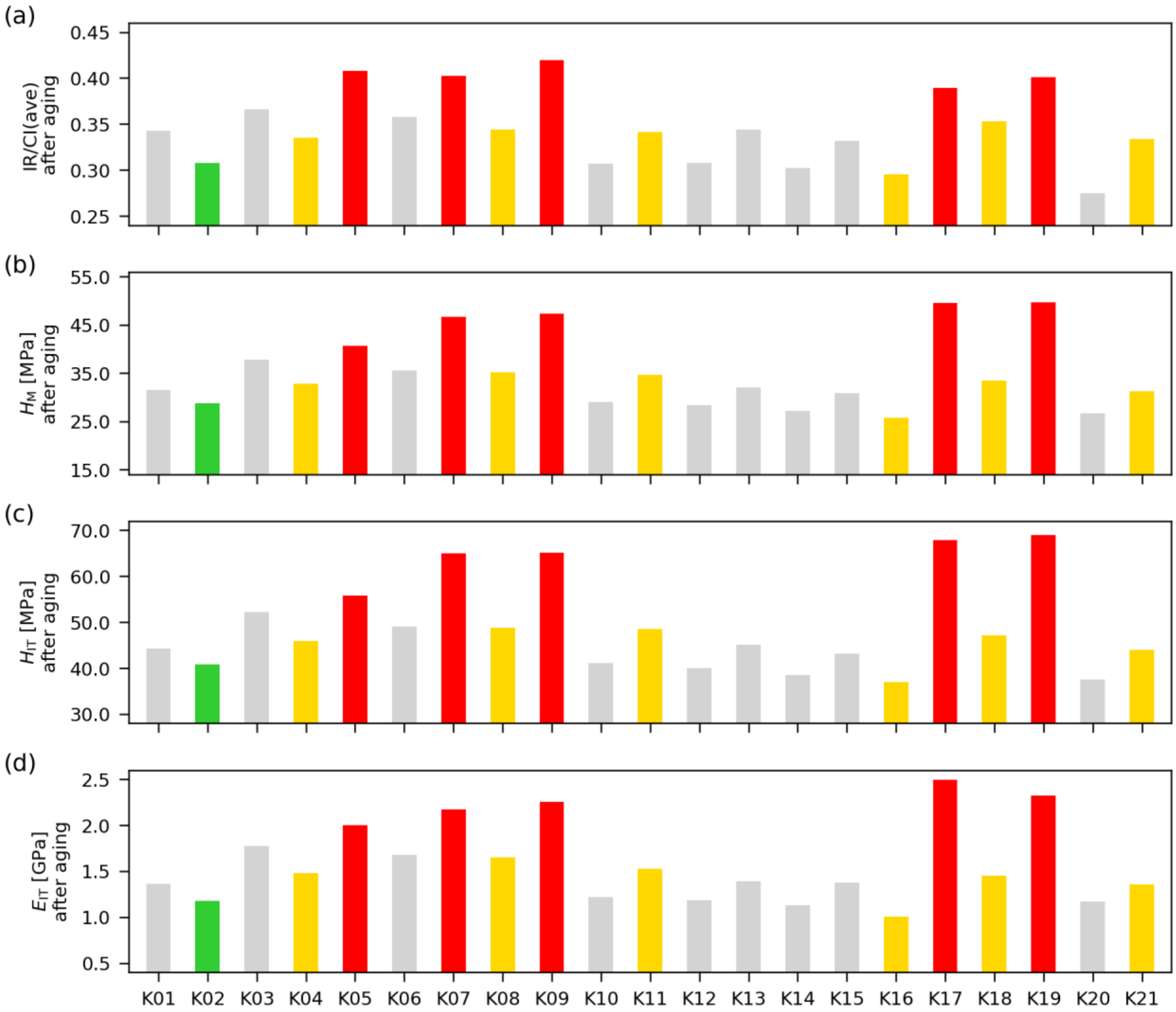

**Figure 8**: Average crystallinity index from IR (a), and Martens hardness (b), indentation hardness (c), and indentation modulus (d) from MHI. All experimental data for IR/OI(ave), MHI/$H_M$, MHI/$H_{IT}$ and MHI/$E_{IT}$ come from the central region of the sample. The column colors are the same as in Fig. 1, which shows the samples before the accelerated aging: red = samples with residual radicals and no stabilizer, orange = samples with residual radicals and a stabilizer, green = sample without residual radicals and with a stabilizer, gray = all other samples (without residual radicals and a stabilizer). The standard deviations did not exceed 5% of the measured values.

# 14 Figure 9

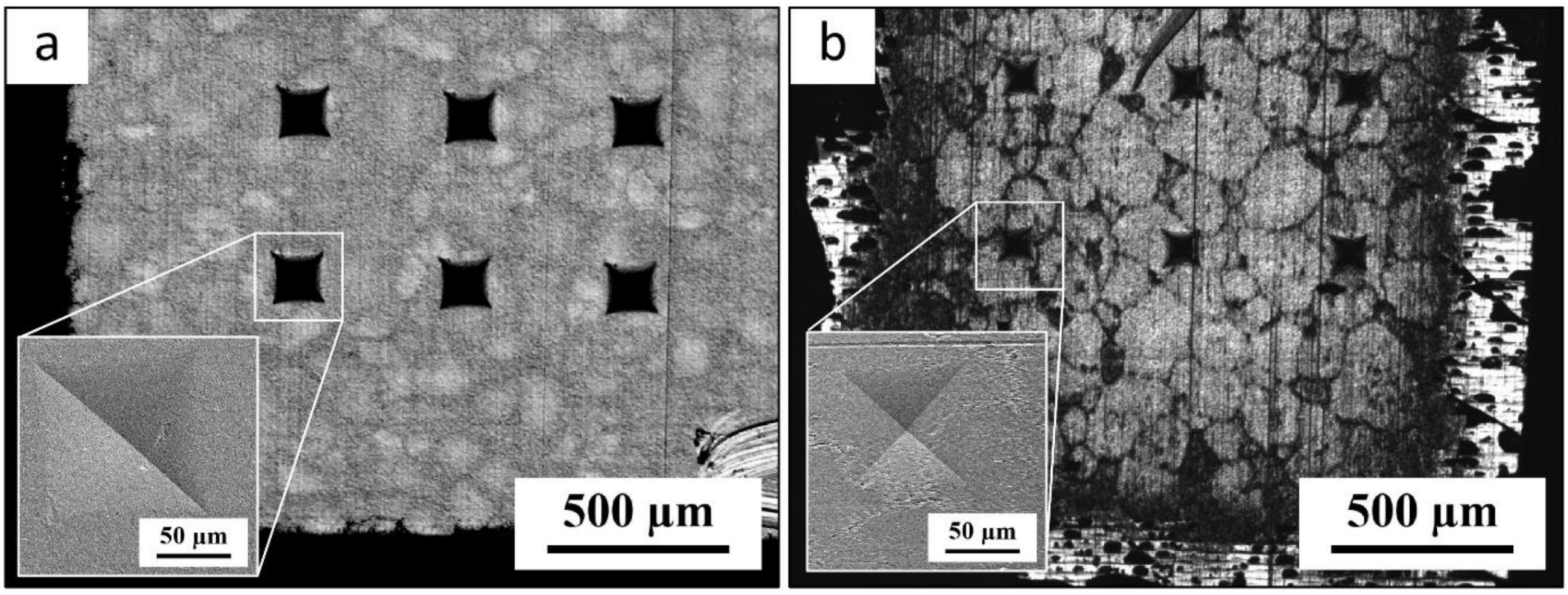

**Figure 9**: Reflected light micrographs showing cut surfaces for two typical UHMWPE formulations: (a) sample K16 with no residual radicals and low oxidative degradation and (b) sample K19 with residual radicals and heavy oxidative degradation. More details about the samples can be found in Table 1. Both cut surfaces contain six imprints of the Vickers tip from the microindentation testing. The insets are SEM micrographs showing the detail of a selected imprint.

# 15 Figure 10

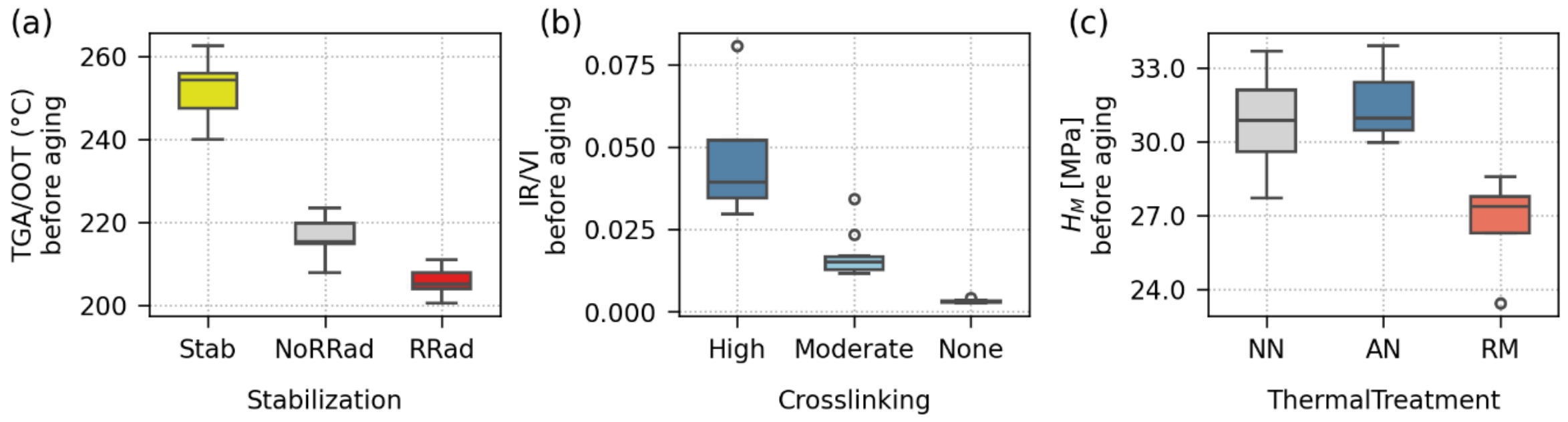

**Figure 10**: Boxplots showing the differences among the selected groups of UHMWPE samples *before the accelerated aging*. Figure 10a displays oxidation onset temperatures from TGA for samples with different types of stabilization: samples with a biocompatible stabilizer (yellow box), samples without stabilizer and without residual radicals (gray box), and samples without stabilizer and with residual radicals (red box). Figure 10b displays trans-vinylene indexes from IR for samples with high degree of crosslinking (radiation doses above 100 kGy; dark blue box), moderate degree of crosslinking (nonzero doses below 100 kGy; light blue box), and samples without crosslinking and/or sterilization by radiation (gray box). Figure 10c displays Martens hardness from MHI for samples without thermal treatment (gray box), samples modified by annealing (blue box), and samples modified by remelting (orange box).

# 16 Figure 11

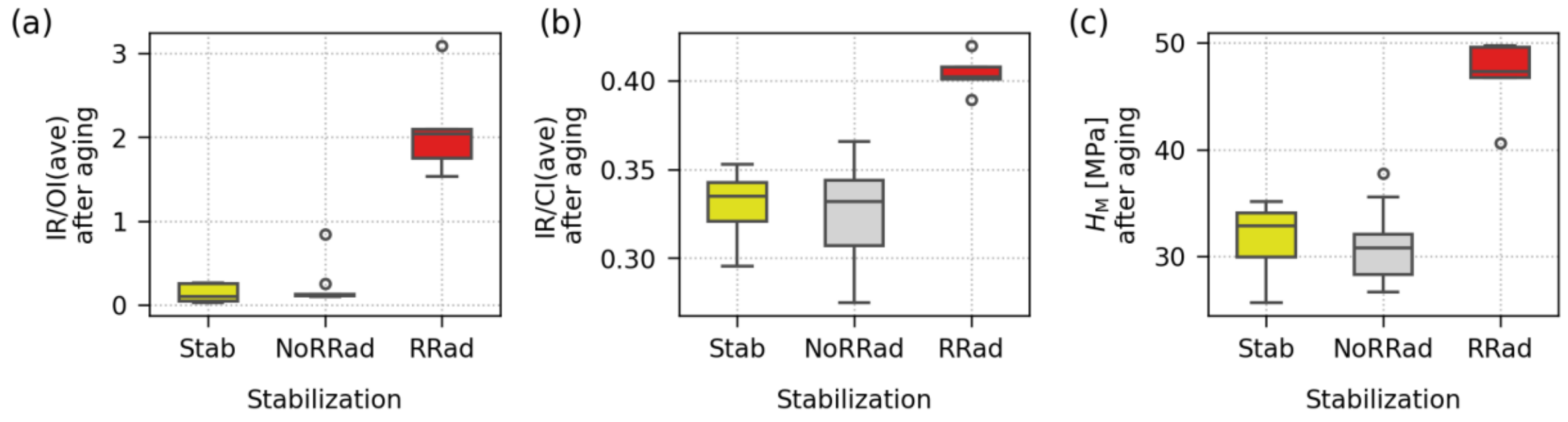

**Figure 11**: Boxplots showing the differences among the selected groups of UHMWPE samples *after the accelerated aging*. All three plots compare the same subsets of samples, grouped according to the type of stabilization, which was the decisive parameter influencing the properties of aged UHMWPE samples: yellow boxes = samples with a biocompatible stabilizer, gray boxes = samples without a stabilizer and without residual radicals, and red boxes = samples without stabilizer and with residual radicals. The figures show the impact of stabilization on: (a) average oxidation index from IR microspectroscopy, (b) average crystallinity from IR microspectroscopy, and (c) Martens hardness from microindentation testing.

# 17 Figure 12

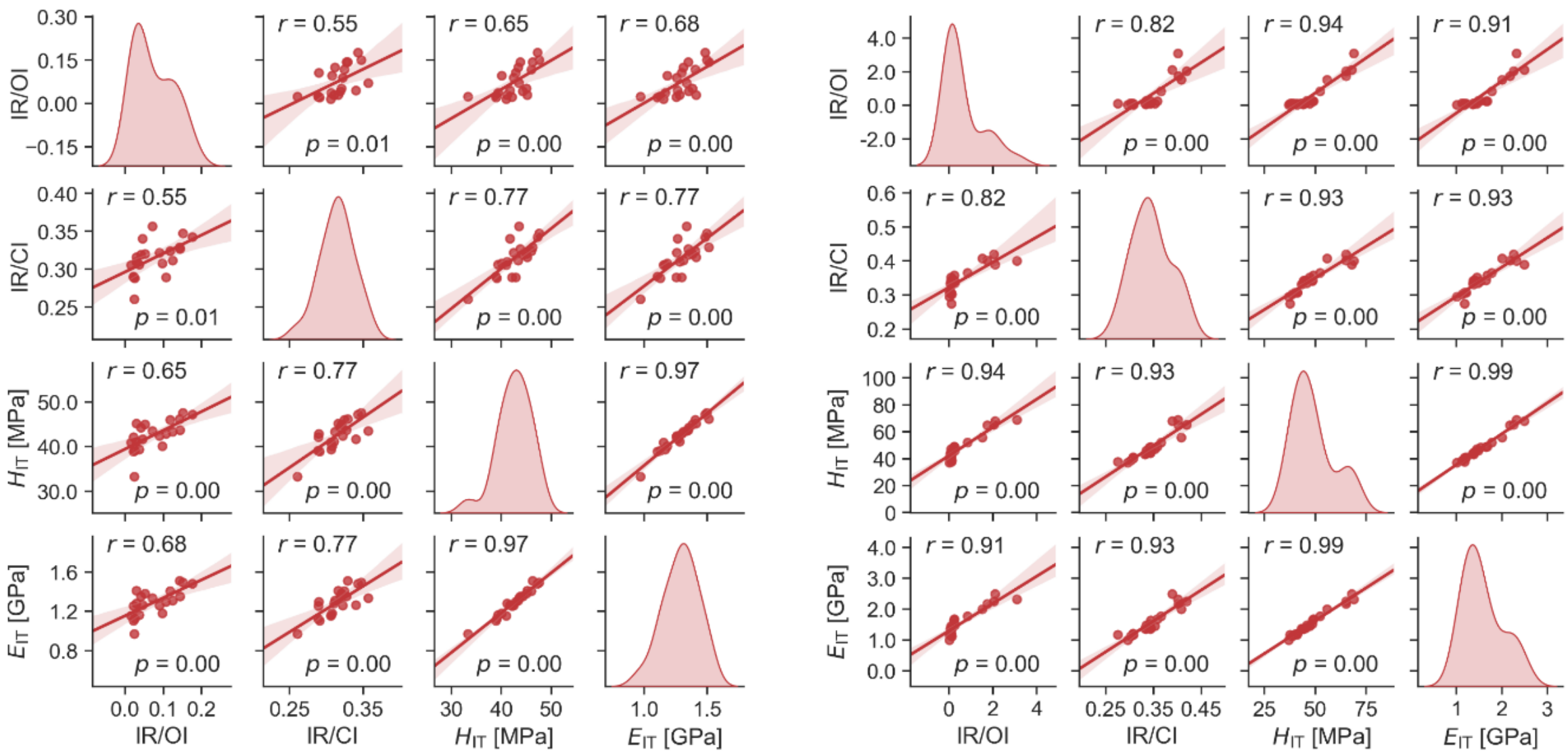

(a) Structure-property correlations BEFORE aging

(b) Structure-property correlations AFTER aging

**Figure 12**: Scatterplot matrix graph showing correlations between oxidation (IR/OI), crystallinity (IR/CI), and micromechanical properties ($H_{IT}$ and $E_{IT}$) for all UHMWPE samples before (a) and after (b) the accelerated aging. Diagonal elements of the scatterplot matrix graph show distribution of the measured quantities, whereas off-diagonal elements show correlations between each pair of quantities. The translucent bands around the regression lines represent 95% confidence interval of the regression estimate. All off-diagonal plots show the values of Pearson's correlation coefficient *r* and *p*-values in the upper left corner and lower right corner, respectively. The high *r* and low *p* values indicate strong and statistically significant correlations; more details about *r* coefficients and *p* values can be found in section 2.5.

# Supplementary information file

## Comparison of various UHMWPE formulations from contemporary total knee replacements before and after accelerated aging

Petr Fulin[1], Veronika Gajdosova[2], Ivana Sloufova[3], Jiri Hodan[2], David Pokorny[1], Miroslav Slouf[2]*

[1] 1st Department of Orthopaedics, First Faculty of Medicine of Charles University and Motol University Hospital, V Uvalu 84, 15006 Prague, Czech Republic

[2] Institute of Macromolecular Chemistry, Czech Academy of Sciences, Heyrovskeho nam. 2, 16206 Praha 6, Czech Republic

[3] Charles University, Faculty of Science, Department of Physical and Macromolecular Chemistry, Hlavova 2030, 128 40 Prague 2, Czech Republic

* Corresponding author e-mail: slouf@imc.cas.cz

## Contents

# 1 IR spectra of UHMWPE samples after *in vivo* and *in vitro* aging

The main text (sections 2.3 and 3.3.1) describes our new protocol of accelerated aging, which was applied to all UHMWPE samples of this contribution. The protocol is based on the submerging of UHMWPE samples into $H_2O_2$ aqueous solution at 70 °C for 30 days. Figure S1 compares the IR spectra of a strongly oxidized UHMWPE sample after 11 years of *in vivo* aging (Figs. S1a–b) with the IR spectra of a strongly oxidized UHMWPE sample aged *in vitro* using our protocol (Figs. S1c–d). The decomposition of the strongest oxidation peak around 1720 $cm^{-1}$ (Figs. 1a, c) documents that our accelerated aging does not produce non-physiological aldehyde oxidation peaks at 1732 $cm^{-1}$ [Kurtz 2009]. The images showing the whole IR spectra (Figs. 1c, d) illustrate that both in vivo and in vitro aged samples contain the same oxidation-related peaks around 940, 1405, 1720, and 3530 $cm^{-1}$. We conclude that our accelerated aging protocol is faster then previously published methods (30 days) and yields IR spectra fully comparable to those of *in vivo* aged UHMWPE inserts.

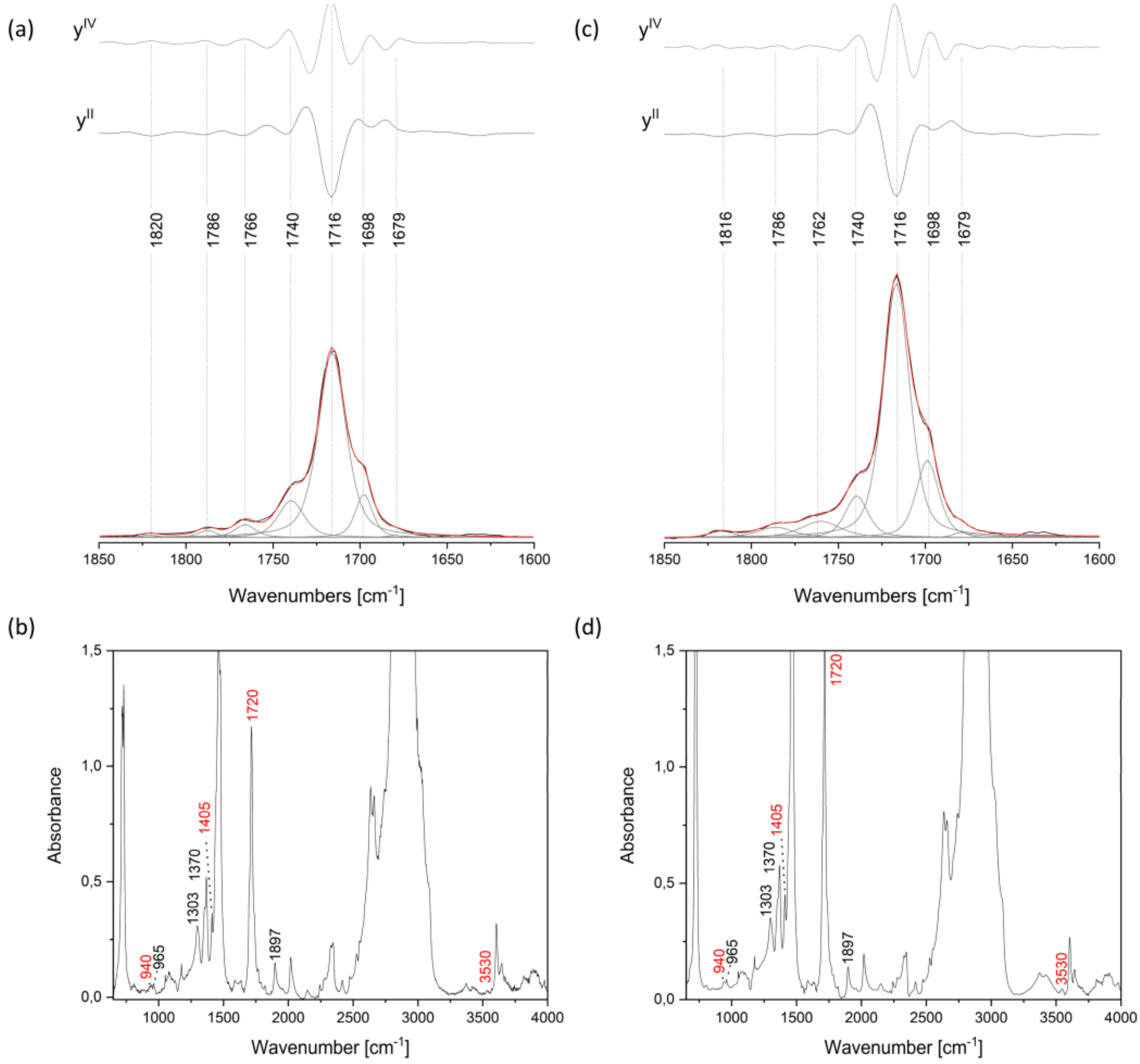

**Figure S1:** IR spectra of UHMWPE samples: (a, b) after 11 years of *in vivo* aging and (c, d) after 30 days of *in vitro* accelerated aging in $H_2O_2$ using our new protocol. Upper row (a, c): decomposition of the main oxidation peak around 1720 cm-1. Lower row (b, d): Whole UHMWPE spectra with the main oxidation-related peaks marked with red font.

# 2 ESR results: residual radicals in UHMWPE samples

Figure S2 summarizes the results of ESR measurements. The ESR measurements were performed for all samples at the end of all experiments in order to verify that the residual radicals are able to survive in the modified polymers. The UHMWPE liners were bought from the manufacturers in year 2020, and the ESR measurements were performed in year 2024, i.e. after four years. Figure S2 documents that low, but non-negligible concentration of residual radicals was present in all samples, which fell in one of the following categories:

- Samples sterilized with ionizing radiation, such as gamma rays or electron beam
- Samples crosslinked with ionizing radiation, which were not remelted and stabilized
- Samples crosslinked with ionizing radiation, which were not remelted, but contained a biocompatible stabilizer (vitamin E or Covernox®)

Figure S2 shows ESR curves for all samples with residual radicals (samples K04, K05, K07, K08, K09, K11, K16, K18, K19, K21) and one representative curve of the sample without residual radicals (sample K02; all samples without residual radicals exhibited the same featureless ESR curve). The full list of samples is in Table 1 in the main text. The non-centrosymmetric shape of the curves results from the fact that the samples contained a mixture of radicals with various g-factors. The radical concentrations were estimated from the double integral of the spectra, which is proportional to the number of detected spins. Complete results of all measurements in the form of Excel table are in the Supplementary information file #2.

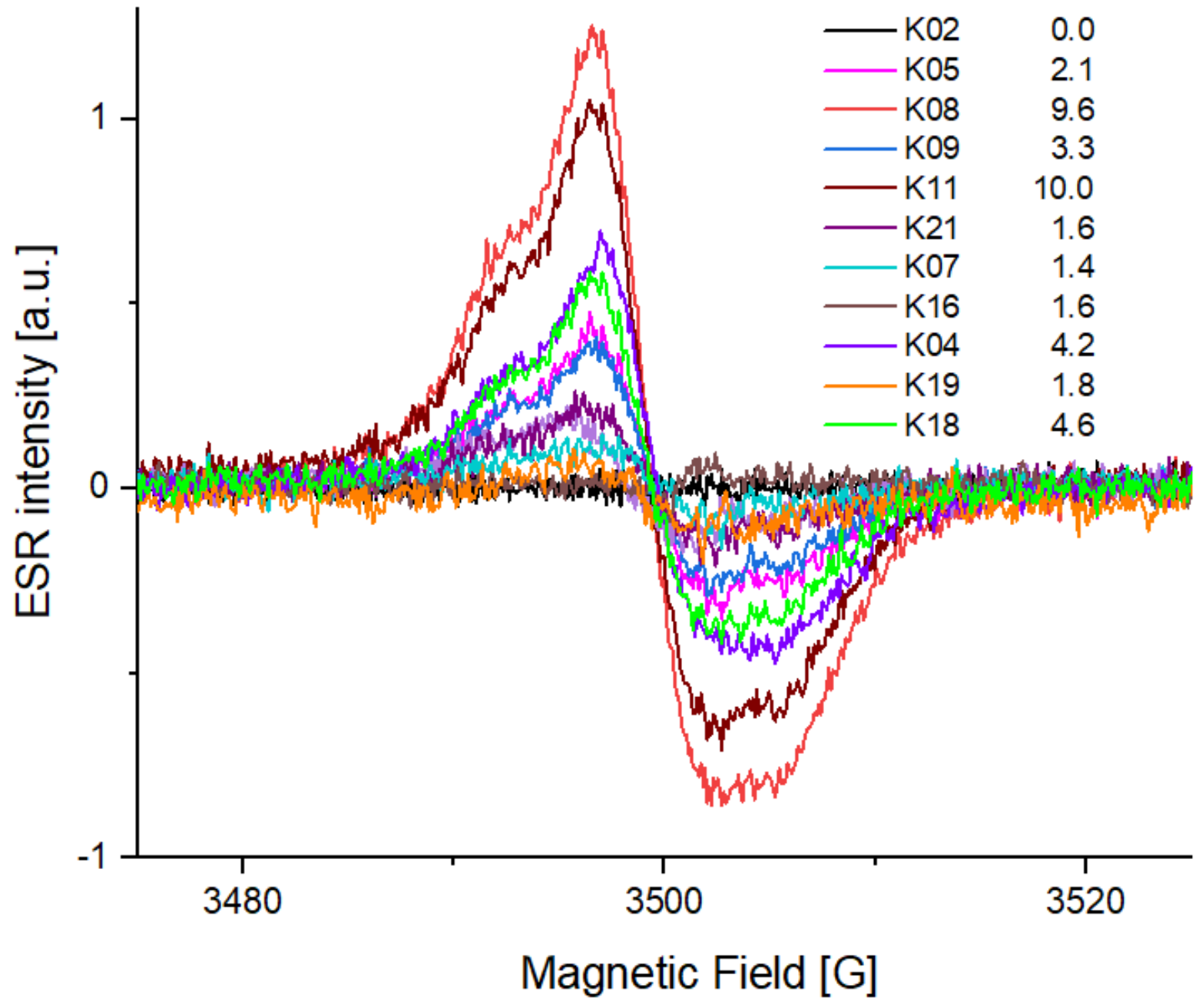

**Figure S2:** ESR spectra of all samples containing residual radicals. The concentrations of the free radicals (numbers shown in the legend next each sample ID) are on relative scale, normalized to 10 and the approximate calibration suggested that they are in the order of $10^{-9}$ spin/g.

# 3 Surface morphology and properties of aged UHMWPEs

Micromechanical properties of UHMWPE liners after the accelerated aging could not be measured reliably and reproducibly from the polymer surfaces due to a significant surface damage. Roughness and/or microcracks were observed on all surfaces, regardless of their modification, as documented by both light microscopy (Fig. S3) and scanning electron microscopy (Fig. S4). These damaged surfaces are typical of $H_2O_2$ aging, but explanted liners suffer from similar problems.

The results of IR and MHI measurements from the surface of the UHMWPE samples are summarized in Fig. S5. The data are rather scattered and the differences among sample groups (represented by red, orange, green and gray columns in Fig. S5) are hardly visible (the column colors are described in the figure legend and are the same as in Fig. 8 in the main text). If the properties were measured from the central region, the differences were clear as documented in Figs. 8 and 9 of the main text. Complete results of all measurements in the form of Excel table are in the Supplementary information file #2.

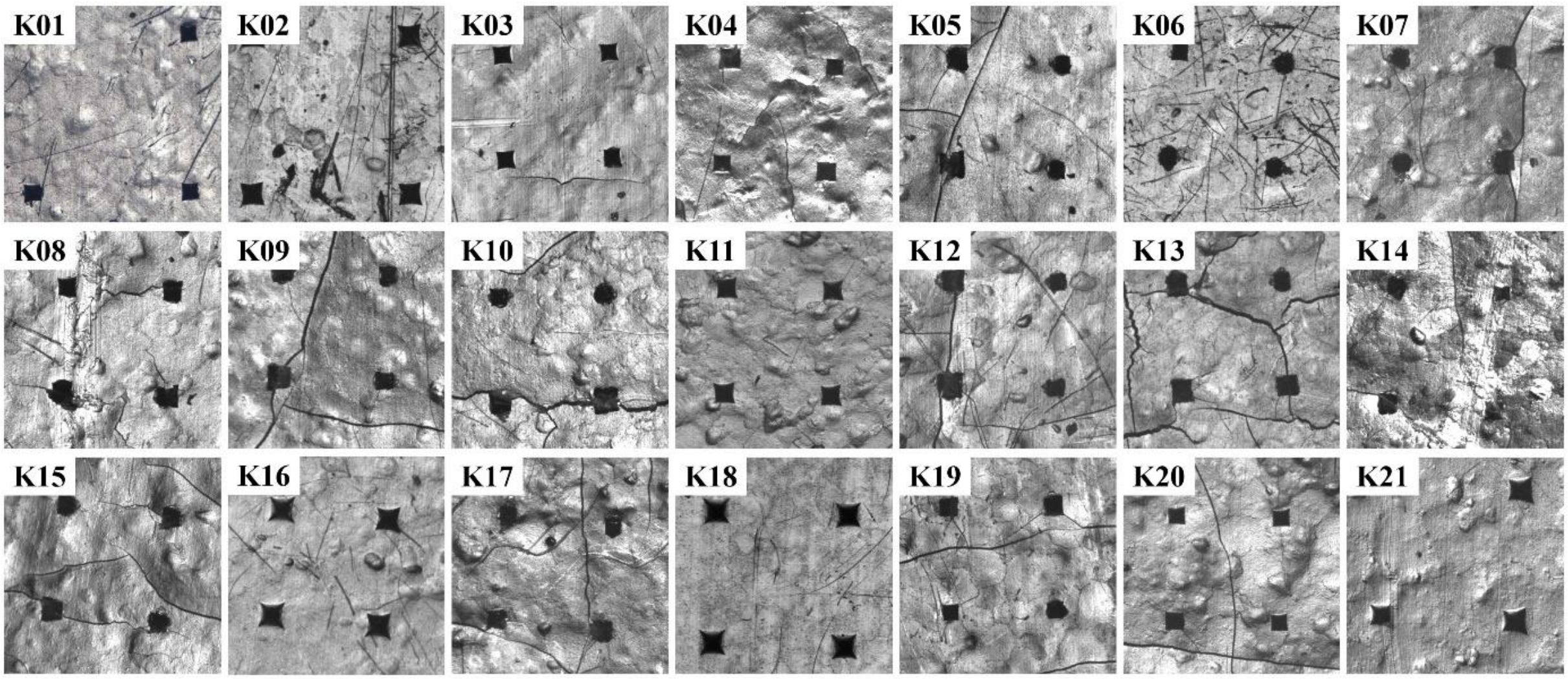

**Figure S3:** Light micrographs of UHMWPE samples surface after 30 days of accelerating aging. Each surface contains four imprints of the indenter tip from MHI measurements. Micrographs are given without scalebar, RWI of all images is ca 830 µm.

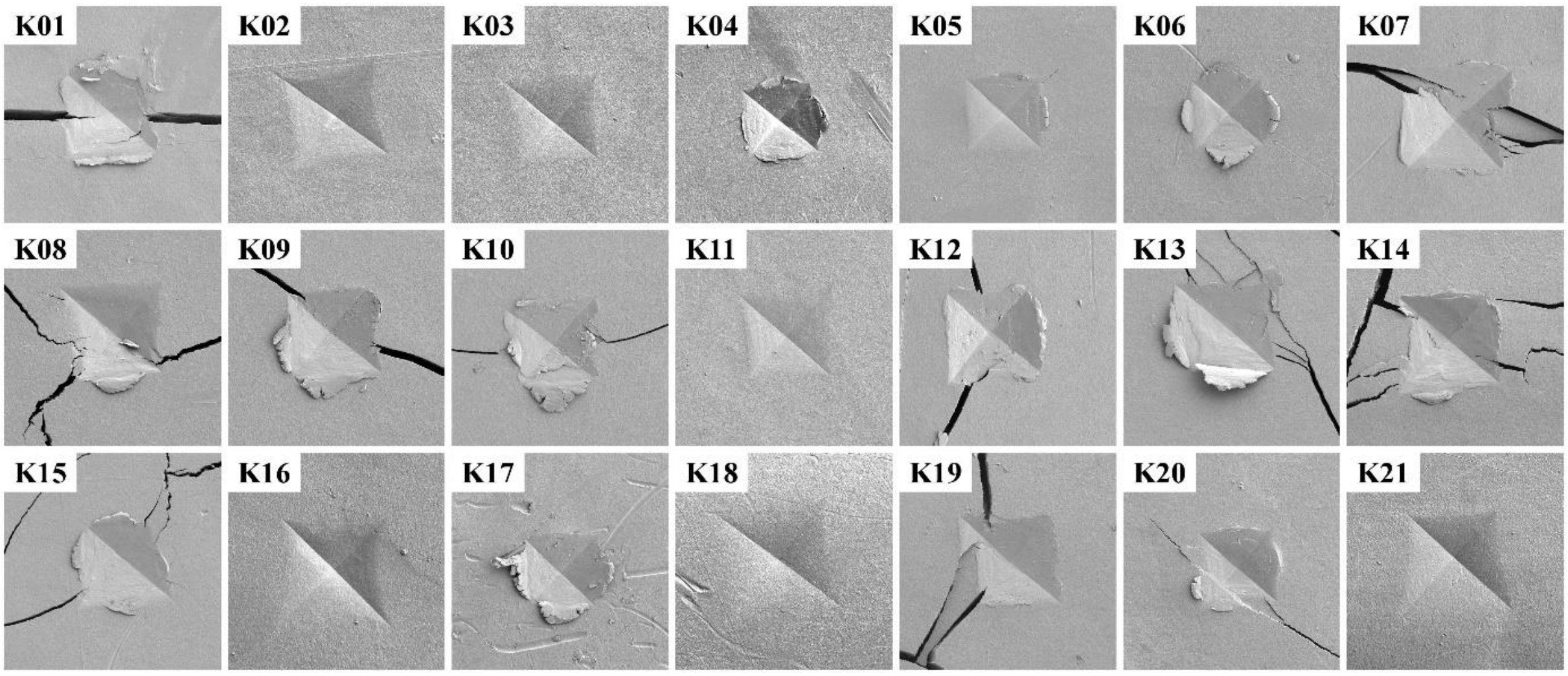

**Figure S4:** SEM micrographs of UHMWPE samples surface after 30 days of accelerating aging. Each surface contains an imprint of the indenter tip from MHI measurements. Micrographs are given without scalebar, RWI of all samples is ca 240 µm.

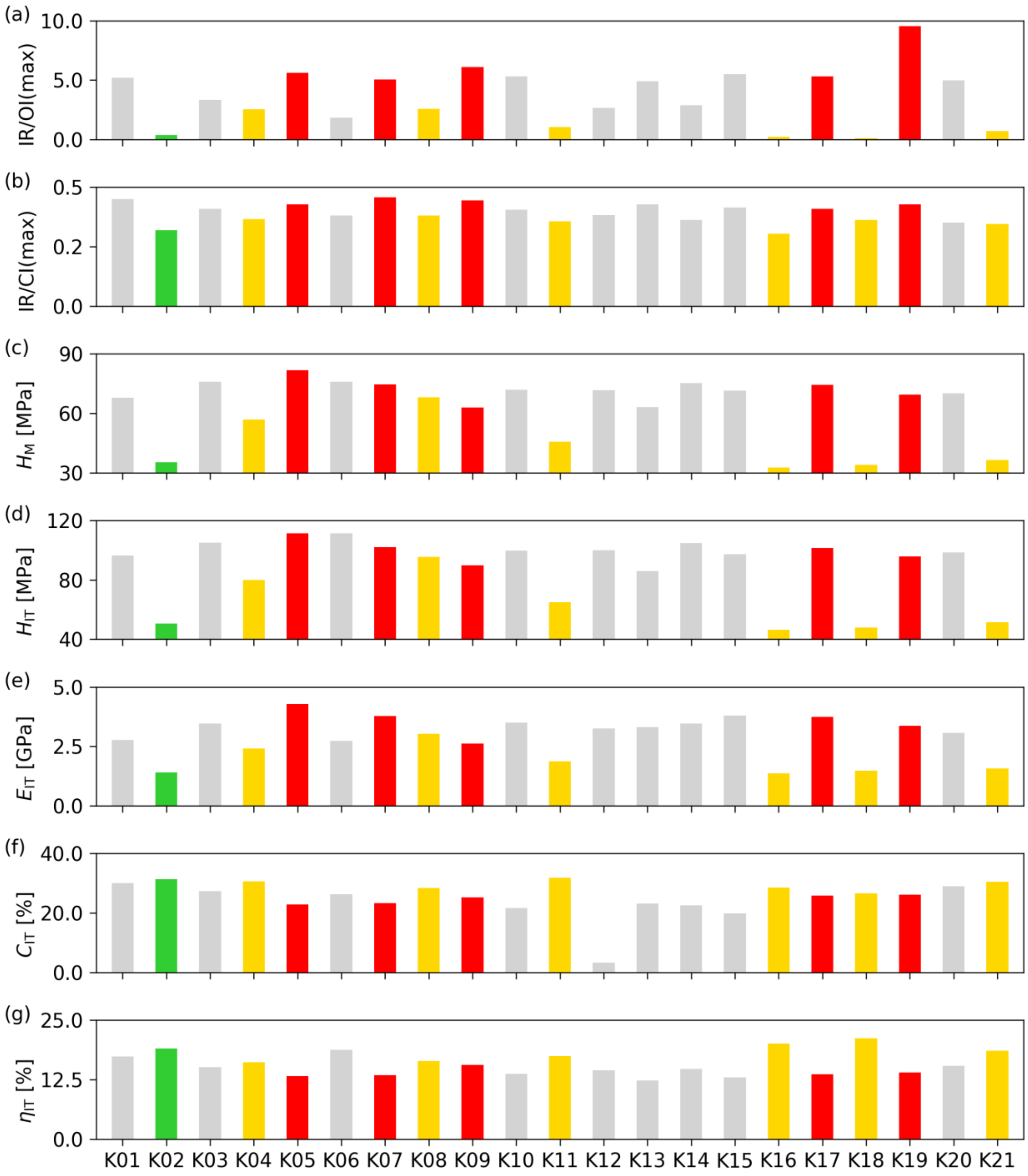

**Figure S5:** Maximal oxidation index (a) and maximal crystallinity index (b) from IR, Martens hardness (c), indentation hardness (d), indentation modulus (e), indentation creep (f) and elastic part of the indentation work (g) from MHI. All experimental data for come from the aged surface of the sample. The column colors are the same as in Fig. 8 in the main text, which shows the properties of samples

after the accelerated aging measured from the central region: red = samples with residual radicals and no stabilizer, orange = samples with residual radicals and a stabilizer, green = sample without residual radicals and with a stabilizer, gray = all other samples (without residual radicals and a stabilizer).

# 4 Complete results of microindentation measurements

Figure S6 displays correlation between oxidation index (IR/OI), crystallinity index (IR/CI) and all micromechanical properties ($E_{IT}$, $H_{IT}$, $H_M$, $C_{IT}$, and $\eta_{IT}$) from MHI measurements. The stiffness-related properties ($E_{IT}$, $H_{IT}$, and $H_M$) correlate strongly with OI and CI, while the viscosity-related properties ($C_{IT}$ and $\eta_{IT}$) show weaker correlations documenting the increase in plasticity with crystallinity. Complete results of all measurements in the form of Excel table are in the Supplementary information file #2.

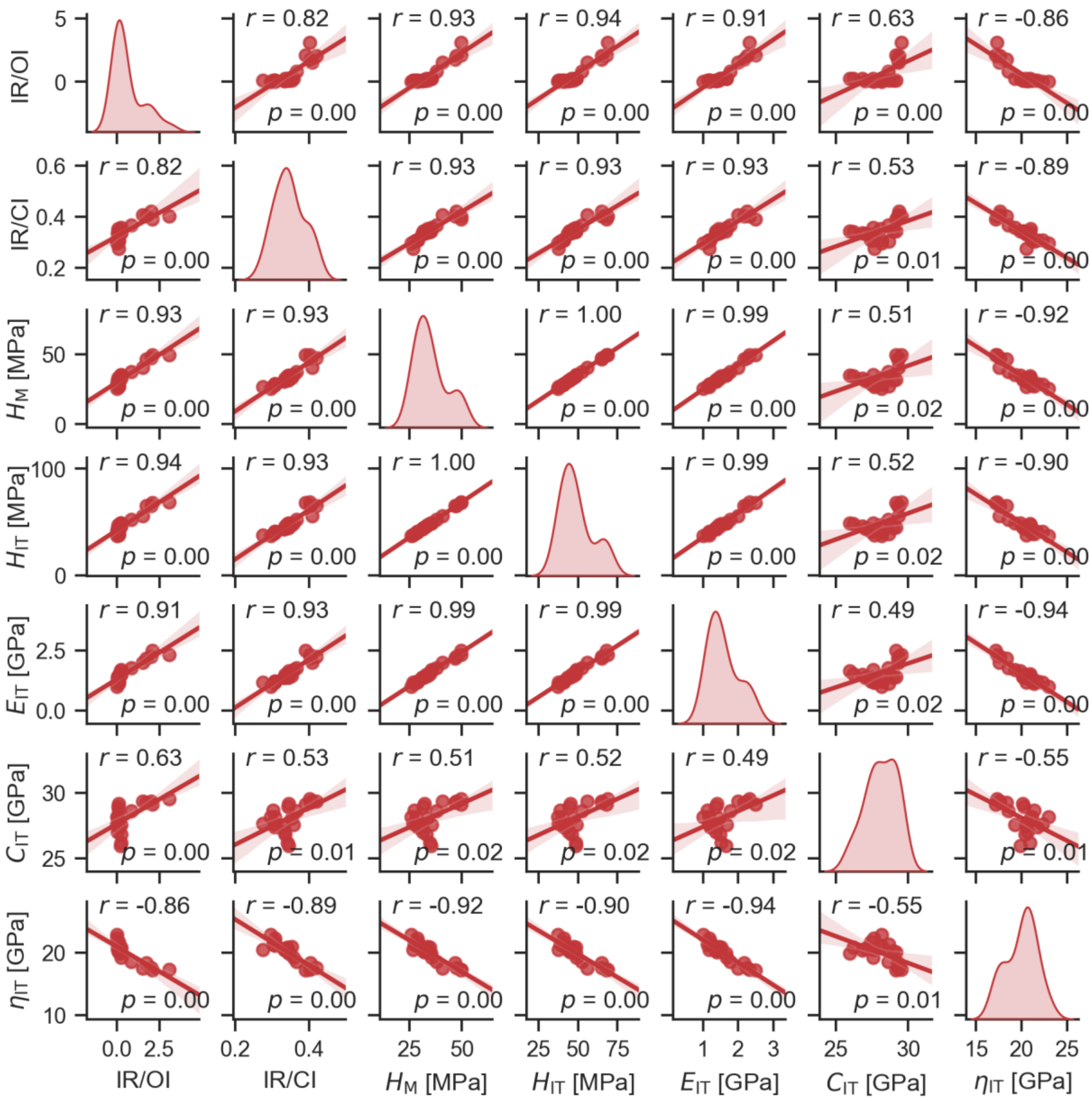

**Figure S6:** Scatterplot matrix graph showing correlations between oxidation (IR/OI), crystallinity (IR/CI), and all micromechanical properties ($H_M$, $H_{IT}$, $E_{IT}$, $C_{IT}$ and $\eta_{IT}$) for all UHMWPE samples after the accelerated aging. Diagonal elements of the scatterplot matrix graph show distribution of the measured quantities, whereas off-diagonal elements show correlations between each pair of quantities. The translucent bands around the regression lines represent 95% confidence interval of the regression estimate. All off-diagonal plots show the values of Pearson's correlation coefficient *r* and *p*-

values in the upper left corner and lower right corner, respectively. The high *r* and low *p* values indicate strong and statistically significant correlations; more details about *r* coefficients and *p* values can be found in section 2.5 of the main text.

# 5 Complete results of all measurements

Complete results of all measurements in the form of Excel table are in the Supplementary information file #2.

---

suppl-info-file_no2.xlsx on request